\documentclass[aip, 11pt, notitlepage, reprint, amsmath,nonatbib]{revtex4-2}
\usepackage{graphicx}
\usepackage{amsmath}
\usepackage[normalem]{ulem}
\usepackage{hyperref}
\usepackage{subcaption}
\usepackage{multirow}
\usepackage[utf8]{inputenc}
\renewcommand{\selectlanguage}[1]{}

\usepackage[T1]{fontenc}
\usepackage{booktabs}
\usepackage{amsfonts}
\usepackage{nicefrac}
\usepackage{microtype}
\usepackage{xcolor}
\newcommand{\Tau}{\tau}
\usepackage[labelfont=bf]{caption}
\newcommand{\anisoap}{\texttt{AniSOAP}}
\begin{document}
\title{A Generalized Approach for Incorporating Geometry and Directionality into Coarse-Grained Machine-Learned Potentials}

\author{Arthur Y. Lin}
\affiliation{Department of Chemical and Biological Engineering, University of Wisconsin - Madison, Madison, WI, USA}
\author{Tejas Dahiya}
\affiliation{Department of Computer Science, University of Wisconsin - Madison, Madison, WI, USA}
\author{Rose K. Cersonsky}
\email{rose.cersonsky@wisc.edu}
\affiliation{Department of Chemical and Biological Engineering, University of Wisconsin - Madison, Madison, WI, USA}
\affiliation{Department of Materials Science and Engineering, University of Wisconsin - Madison, Madison, WI, USA}
\affiliation{Data Science Institute, University of Wisconsin - Madison, Madison, WI, USA}

\begin{abstract}
\end{abstract}

\maketitle

\section{Introduction}
Machine-learned interatomic potentials (MLIPs) have dramatically expanded the scope of molecular simulation, enabling the prediction of atomic energies and forces with near ab initio accuracy at costs compatible with large-scale molecular dynamics.\cite{behler_generalized_2007, unke_machine_2021, batatia_foundation_2025}
As these models continue to mature, the principal obstacle to computational discovery is increasingly shifting from the accuracy of atomistic interactions to the challenge of accessing the larger length and time scales associated with collective behavior, self-assembly, and phase transitions. Coarse-graining (CG), in which groups of atoms are mapped onto effective particles with reduced degrees of freedom, remains one of the most powerful approaches for bridging this gap.\cite{noid_rigorous_2024, noid_multiscale_2008, zadok_coarse-grained_2018, gorlich_mapping_2026, wang_machine_2019}

For any MLIP, its capabilities are determined jointly by the quality of the underlying data, the choice of representation $X$, and the architecture $f(X)$, where $X$ varies from simple forms (e.g., distance matrices or graphs) to more complex descriptors (e.g., Smooth Overlap of Atomic Positions [SOAP]\cite{bartok_representing_2013} or Behler-Parinello symmetry functions\cite{behler_neural_2011}). 
A representation is \emph{complete} if distinct data objects map to distinct data points; in the atomistic MLIP community there has been considerable discussion on what constitutes a \emph{complete} descriptor, as this has direct implications to the ceiling of predictive accuracy in subsequent models.\cite{anstine_machine_2023, pozdnyakov_incompleteness_2022, pozdnyakov_smooth_2023, pozdnyakov_incompleteness_2020}. 
Simply put, while expressive architectures can approximate complex functions $f(X)$, they cannot recover information that is not encoded in $X$, and degeneracy in data representation can explicitly limit model performance. 

However, unlike atomistic machine-learned potentials, the variables describing a coarse-grained system are not fixed, but instead chosen. 
Historically, much of coarse-graining has therefore been built around isotropic particles.\cite{noid_rigorous_2024}  This choice is attractive for both conceptual and computational reasons. Interactions become functions of intermolecular separation alone, simulation methodologies are well-established, and many successful coarse-grained models have been developed within this framework.\cite{marrink_martini_2007} Such approaches have proven remarkably successful, particularly in biomolecular systems where coarse-graining has enabled simulations inaccessible at atomistic resolution.\cite{marrink_martini_2007}  
Yet there are also many examples where isotropic descriptions struggle to reproduce experimentally observed behavior, particularly in systems whose organization is governed by packing, local structure, or directional interactions.\cite{hosseini_martini_2024,loose_changing_2024}

One possible explanation of these difficulties is that geometry itself carries critical information. Consider two molecular configurations possessing similar intermolecular separations but different relative orientations. Depending on the system, these configurations may exhibit substantially different interaction energies. When orientation is removed from the coarse-grained representation, however, both are represented by the same set of variables $\mathbf{R}$. The resultant coarse-grained configuration is therefore associated not with a single underlying interaction energy, but with a distribution of possible interaction energies. As this distribution broadens, the construction of a coarse-grained potential becomes increasingly difficult, as distinct molecular environments become indistinguishable within the representation.

This observation naturally suggests retaining orientational and geometric information within coarse-grained models. However, doing so raises a separate question: how should such information be represented? While atomistic machine-learned potentials have undergone rapid development over the past decade, the overwhelming majority of descriptor frameworks and architectures were developed for collections of isotropic atoms. Extending these approaches to anisotropic particles requires representations that remain sensitive to particle geometry while preserving the rotational and translational symmetries of the underlying problem.

Several recent approaches have begun to address this challenge through anisotropic extensions to machine-learning architectures.\cite{nguyen_systematic_2022,wilson_anisotropic_2023,loose_coarse-graining_2023} Of particular interest are many-body density expansion methods, which have proven highly successful in atomistic machine learning due to their systematic improvability and clear connection to local structure.\cite{bartok_representing_2013,drautz_atomic_2019,batatia_mace_2022} \anisoap~extended the SOAP formalism to non-spherical particles through anisotropic density expansions, enabling the explicit incorporation of particle shape and orientation into a many-body descriptor,\cite{lin_expanding_2024,lin_anisoap_2025}, and similar efforts expanded the Chebyshev polynomial suite for nanoscale particles.\cite{fakhraei_approximation_2025, fakhraei_approximation_2026, zhang_aniso-chimes_2026}.

In this work, we demonstrate the strategies and subsequent trade-offs for incorporating geometric information into MLIPs. First, we show how incorporating geometric tensors, both through \anisoap~and as node features in MACE, compare to their isotropic analogs as a function of molecular symmetry and objective weighting. Then, we demonstrate how to successfully incorporate geometric information into MACE-style potentials for arbitrary molecular symmetries, in ways that extend to coarse-graining at any resolution, even that of rigid composite bodies typically found in the nanoscale community. We do so for single-bead coarse-grained systems with minimal mapping entropy in order to demonstrate the limitations of potential energy fitting for coarse-graining, but will note that analogous schemes can be applied to learn the potential of mean force. Our coinciding packages, \anisoap\cite{lin_anisoap_2025} and MACE-CG 
are also included open-source, to allow for adoption of this pipeline for new coarse-graining tasks. 

\begin{figure*}
    \centering
    \includegraphics[width=\linewidth]{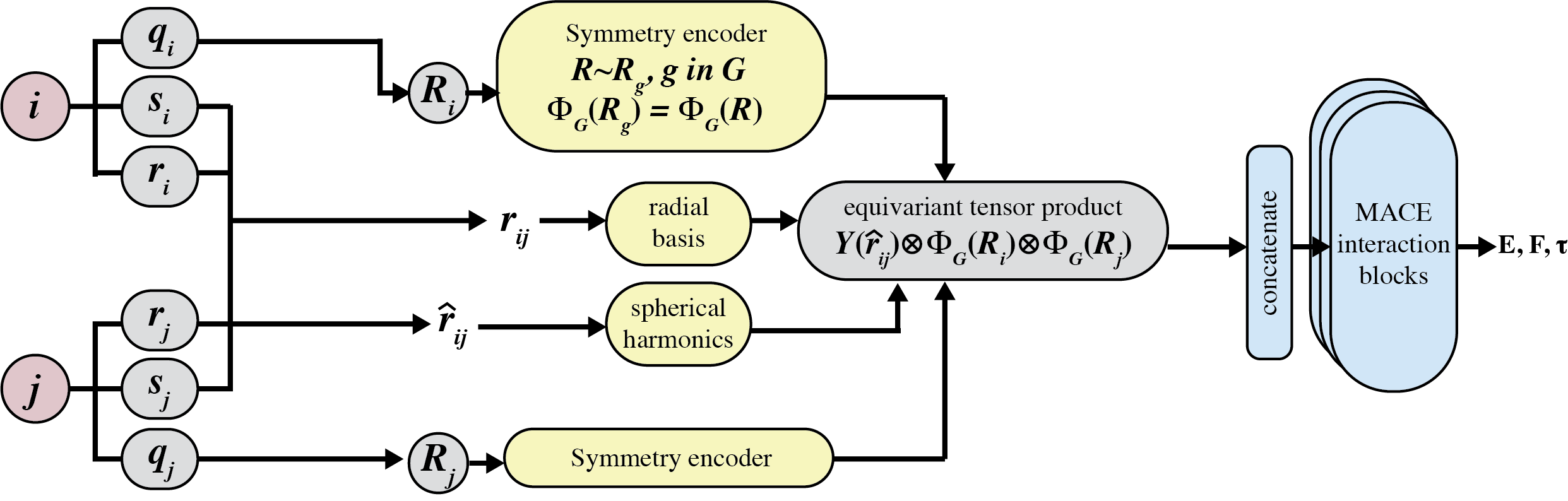}
    \caption{Workflow for incorporating molecular geometry and directionality into MACE-style potentials.}
    \label{fig:mace_aniso}
\end{figure*}
\section{Methods}

\subsection{\anisoap~Descriptors}
Many-body density expansions provide a natural framework for constructing machine-learning representations of local environments. In conventional SOAP-based approaches, each particle is represented by an isotropic Gaussian density centered at its position. \anisoap\cite{lin_anisoap_2025, lin_expanding_2024} extends this formalism to anisotropic particles by replacing isotropic densities with ellipsoidal density fields that explicitly depend on particle geometry and orientation.

Each coarse-grained particle is described by a position $\mathbf{r}_i$, an orientation matrix $R_i$, and geometric parameters corresponding to its principal semiaxes $(a_i,b_i,c_i)$. The particle geometry is encoded through the matrix
\begin{equation}
\Sigma_i
=
R_i
\begin{pmatrix}
a_i^2 & 0 & 0 \\
0 & b_i^2 & 0 \\
0 & 0 & c_i^2
\end{pmatrix}
R_i^{T}.
\end{equation}
The local density surrounding particle $i$ is then written as
\begin{equation}
\rho_i(\mathbf r)
=
\sum_j
\exp
\left[
-\frac{1}{2}
(\mathbf r-\mathbf r_{ij})^{T}
\Sigma_j^{-1}
(\mathbf r-\mathbf r_{ij})
\right],
\end{equation}
where $\mathbf r_{ij}=\mathbf r_j-\mathbf r_i$ denotes the position of neighboring particle $j$ relative to particle $i$.
Following the SOAP formalism, the density is expanded in a basis of radial functions and spherical harmonics,
\begin{equation}
\rho_i(\mathbf r)
=
\sum_{nlm}
c_{nlm}^{(i)}
g_n(r)
Y_{lm}(\hat{\mathbf r}),
\end{equation}
producing a set of rotationally equivariant coefficients
$c_{nlm}^{(i)}$.
Rotationally invariant features are subsequently obtained through contractions of these coefficients. Throughout this work, we employ the \anisoap~power spectrum,
\begin{equation}
p_{nn'l}^{(i)}
=
\sum_m
c_{nlm}^{(i)}
c_{n'lm}^{(i)*},
\end{equation}
which forms the descriptor vector $\mathbf x_i$ used for regression. By construction, the descriptor is invariant to global rotations and translations while remaining sensitive to particle geometry and local orientational environments.

We can consider linear energy models acting on the \anisoap~power spectrum written as
\begin{equation}
E
=
\sum_i
\mathbf{x}_i^T \mathbf w
\end{equation}
where $\mathbf x_i$ denotes the \anisoap~descriptor associated with particle $i$ and $\mathbf w$ is a learned coefficient vector.

\subsection{Incorporating Rigid-Body Orientation into MACE}
\label{sec:rigid_mace}

As a complementary approach to \anisoap, we extended the MACE equivariant
message-passing framework to describe coarse-grained particles with explicit
rigid-body orientation. The central objective was to preserve the existing
MACE treatment of translational geometry while augmenting each intermolecular
edge with orientational information. In this construction, the molecular
symmetry is incorporated into the representation of the body orientation
before that representation enters the MACE interaction network.

\subsubsection{Standard MACE edge geometry}

For a pair of coarse-grained particles $i$ and $j$, conventional MACE begins
from the relative displacement
\begin{equation}
    \mathbf r_{ij}
    =
    \mathbf r_j-\mathbf r_i,
\end{equation}
which is decomposed into the scalar separation
\begin{equation}
    r_{ij} = \lVert \mathbf r_{ij} \rVert
\end{equation}
and the unit edge direction
\begin{equation}
    \hat{\mathbf r}_{ij}
    =
    \frac{\mathbf r_{ij}}{r_{ij}}.
\end{equation}
The distance is expanded in a radial basis, while the angular dependence is
represented by spherical harmonics,
\begin{equation}
    Y_{\ell m}(\hat{\mathbf r}_{ij}).
\end{equation}
Under a global rotation $S\in SO(3)$, these functions transform according to
the $\ell^\text{th}$ irreducible representation,
\begin{equation}
    Y_{\ell}(S\hat{\mathbf r}_{ij})
    =
    D^{(\ell)}(S)
    Y_{\ell}(\hat{\mathbf r}_{ij}),
\end{equation}
which allows the edge geometry to be coupled to node features through the
Clebsch--Gordan tensor products used throughout MACE.

For an atomistic system, the combination of particle identity and relative
positions is sufficient to define the local geometry. This is no longer true
when an entire molecule is mapped to a single coarse-grained site. Two
molecular configurations can possess identical center-of-mass positions,
and therefore identical $r_{ij}$ and
$Y_{\ell m}(\hat{\mathbf r}_{ij})$, while differing substantially in
molecular orientation and interaction energy. A single-site rigid-body model
therefore requires an additional orientational degree of freedom.

\subsubsection{Rigid-body state and orientation}

Each coarse-grained molecule is represented by
\begin{equation}
    \left(
        \mathbf r_i,
        R_i,
        s_i
    \right),
\end{equation}
where $\mathbf r_i$ is the molecular center-of-mass position,
$R_i\in SO(3)$ is a rotation matrix mapping a fixed molecular body frame to
the simulation frame (we will call this the ``simulation-frame'' rotation), and $s_i$ denotes molecular species. In practice,
orientations are stored as unit quaternions and converted to rotation
matrices before constructing equivariant features.

Writing the columns of $R_i$ as the simulation-frame body axes,
\begin{equation}
    R_i =
    \begin{pmatrix}
        \vert & \vert & \vert \\
        \mathbf a_i & \mathbf b_i & \mathbf c_i \\
        \vert & \vert & \vert
    \end{pmatrix},
\end{equation}
a global rotation $S$ acts as
\begin{equation}
    \mathbf r_i \rightarrow S\mathbf r_i,
    \qquad
    R_i \rightarrow S R_i.
\end{equation}
The body axes therefore transform as ordinary equivariant vectors and can,
in principle, be coupled directly to the spherical-harmonic edge features
using the same tensor-product machinery already present in MACE.

A full-frame rigid-body representation can consequently be constructed from
the equivariant product
\begin{equation}
    Y(\hat{\mathbf r}_{ij})
    \otimes
    R_i
    \otimes
    R_j,
    \label{eq:full-frame-rigid-pair}
\end{equation}
where the three columns of each rotation matrix are treated as $\ell=1$
features. Clebsch--Gordan tensor products decompose this product into
irreducible $SO(3)$ channels, which are subsequently projected into a chosen
set of edge irreducible representations.

Although Eq.~\ref{eq:full-frame-rigid-pair} preserves complete body-frame
information, it contains a redundancy whenever the molecule possesses a
nontrivial rotational point-group symmetry.

\subsubsection{Molecular symmetry and the orientation quotient}

Let $G\subset SO(3)$ denote the proper rotational symmetry group of a
molecule. For any $g\in G$, the two body-frame orientations
\begin{equation}
    R_i
    \qquad\mathrm{and}\qquad
    R_i g
\end{equation}
describe the same physical molecular configuration. The physical
orientational degree of freedom therefore belongs to the quotient space
\begin{equation}
    [R_i]\in SO(3)/G.
\end{equation}
An orientational feature map $\Phi_G(R)$ should accordingly satisfy two
distinct transformation properties. First, a rotation of the entire physical
system should act equivariantly,
\begin{equation}
    \Phi_G(SR)
    =
    D(S)\Phi_G(R),
    \label{eq:global-equivariance}
\end{equation}
where $D(S)$ denotes the appropriate direct sum of irreducible
representations. Second, a molecular symmetry operation should leave the
feature unchanged,
\begin{equation}
    \Phi_G(Rg)
    =
    \Phi_G(R),
    \qquad
    g\in G.
    \label{eq:molecular-symmetry}
\end{equation}
Equations~\ref{eq:global-equivariance} and
\ref{eq:molecular-symmetry} distinguish a global rotation, which must be
handled equivariantly by the network, from a body-frame symmetry operation,
which represents no physical change in the molecular configuration.

More generally, symmetry-adapted body features may be constructed by
projecting each angular-momentum representation onto its
$G$-invariant subspace. For angular momentum $\ell$, the group-averaging
projector is
\begin{equation}
    P_G^{(\ell)}
    =
    \frac{1}{|G|}
    \sum_{g\in G}
    D^{(\ell)}(g).
    \label{eq:group-projector}
\end{equation}
If $B_\ell$ denotes a basis for the image of
$P_G^{(\ell)}$, a symmetry-adapted equivariant feature is
\begin{equation}
    \Phi_\ell(R)
    =
    D^{(\ell)}(R)B_\ell.
    \label{eq:symmetry-adapted-feature}
\end{equation}
Because vectors in $B_\ell$ are invariant under $G$,
\begin{equation}
    \Phi_\ell(Rg)
    =
    D^{(\ell)}(R)
    D^{(\ell)}(g)
    B_\ell
    =
    \Phi_\ell(R),
\end{equation}
while a global rotation gives
\begin{equation}
    \Phi_\ell(SR)
    =
    D^{(\ell)}(S)
    \Phi_\ell(R).
\end{equation}
Thus the molecular symmetry is removed without sacrificing the global
rotational equivariance required by MACE.

\subsubsection{Example: A symmetry-adapted representation for water}

Water provides an important example because its proper rotational symmetry
group is $C_2$. We choose a molecular body frame
$(\mathbf a,\mathbf b,\mathbf c)$ in which $\mathbf a$ is the physical
two-fold symmetry axis. The nontrivial $C_2$ operation is then a rotation by
$\pi$ about $\mathbf a$,
\begin{equation}
    (\mathbf a,\mathbf b,\mathbf c)
    \rightarrow
    (\mathbf a,-\mathbf b,-\mathbf c).
    \label{eq:c2-axis-action}
\end{equation}
The symmetry-adapted water representation used here contains one
$\ell=1$ contribution and one $\ell=2$ contribution.

The $\ell=1$ feature is the oriented $C_2$ axis itself,
\begin{equation}
    \Phi^{(1)}_{C_2}(R_i)
    =
    \mathbf a_i.
\end{equation}
Because $\mathbf a$ is unchanged under Eq.~\ref{eq:c2-axis-action}, this
feature is invariant to the molecular $C_2$ operation while transforming as
an ordinary vector under global rotations.

The remaining transverse orientation is encoded by an $\ell=2$ feature.
In spherical-harmonic notation, the body-frame construction is
\begin{equation}
    \Phi^{(2)}_{C_2}(R_i)
    =
    \frac{
        Y_2(\mathbf b_i)
        -
        Y_2(\mathbf c_i)
    }{\sqrt{2}}.
    \label{eq:c2-l2-feature}
\end{equation}
Equivalently, its Cartesian structure may be understood through the
quadrupolar tensor
\begin{equation}
    Q_i
    =
    \mathbf b_i\mathbf b_i^{T}
    -
    \mathbf c_i\mathbf c_i^{T}.
    \label{eq:c2-cartesian}
\end{equation}
Under the physical $C_2$ operation,
$\mathbf b_i\rightarrow-\mathbf b_i$ and
$\mathbf c_i\rightarrow-\mathbf c_i$, so
\begin{equation}
    Q_i
    \rightarrow
    (-\mathbf b_i)(-\mathbf b_i)^T
    -
    (-\mathbf c_i)(-\mathbf c_i)^T
    =
    Q_i.
\end{equation}
The water body representation can therefore be written schematically as
\begin{equation}
    \Phi_{C_2}(R_i)
    =
    \Phi^{(1)}_{C_2}(R_i)
    \oplus
    \Phi^{(2)}_{C_2}(R_i),
\end{equation}
corresponding to the e3nn irreducible representation content
\begin{equation}
    1o \oplus 2e.
\end{equation}
Importantly, physically equivalent orientations satisfy
\begin{equation}
    \Phi_{C_2}(R_i g)
    =
    \Phi_{C_2}(R_i),
    \qquad
    g\in C_2,
\end{equation}
so the network is not required to learn the molecular $C_2$ symmetry from
training data.

This symmetry treatment differs from using a molecular inertia tensor as the
orientation descriptor. A generic triaxial second-rank tensor is invariant
under $\pi$ rotations about each of its three principal axes and therefore
has the proper rotational stabilizer $D_2$. Such a representation identifies
orientations according to $SO(3)/D_2$. Water, however, requires
$SO(3)/C_2$. Since $C_2$ is a strict subgroup of $D_2$, an inertia-tensor
representation introduces additional, nonphysical orientation
identifications. The symmetry-adapted $C_2$ representation instead retains
the transverse orientational information needed to distinguish these
configurations.

\subsubsection{Rigid-body pair features}

The symmetry-adapted molecular features are introduced into MACE at the
pair-interaction level. For an edge connecting molecules $i$ and $j$, the
rigid-body branch constructs the equivariant tensor product
\begin{equation}
    \mathcal Z_{ij}^{\mathrm{rigid}}
    =
    Y(\hat{\mathbf r}_{ij})
    \otimes
    \Phi_{G_i}(R_i)
    \otimes
    \Phi_{G_j}(R_j).
    \label{eq:rigid-pair-product}
\end{equation}
For the $C_2$ water representation, Eq.~\ref{eq:rigid-pair-product}
therefore contains couplings between the ordinary edge spherical harmonics
and the $\ell=1$ and $\ell=2$ body features of both interacting molecules.
Schematically, these include products of the form
\begin{align}
    Y_\ell(\hat{\mathbf r}_{ij})
    &\otimes
    \Phi_i^{(1)}
    \otimes
    \Phi_j^{(1)}, \\
    Y_\ell(\hat{\mathbf r}_{ij})
    &\otimes
    \Phi_i^{(1)}
    \otimes
    \Phi_j^{(2)}, \\
    Y_\ell(\hat{\mathbf r}_{ij})
    &\otimes
    \Phi_i^{(2)}
    \otimes
    \Phi_j^{(2)}.
\end{align}
The tensor products are decomposed into irreducible representations using
the standard Clebsch--Gordan coupling rules. In component notation, a
representative coupling may be written schematically as
\begin{equation}
    Z^{(L)}_{M}
    =
    \sum_{m,\mu_i,\mu_j}
    C^{LM}_{\ell m,\mu_i,\mu_j}
    Y_{\ell m}(\hat{\mathbf r}_{ij})
    \Phi_{i,\mu_i}
    \Phi_{j,\mu_j},
\end{equation}
where the effective coefficient $C^{LM}_{\ell m,\mu_i,\mu_j}$ denotes the
sequence of Clebsch--Gordan contractions required to couple the three input
irreducible representations to total angular momentum $L$.

Because each factor in Eq.~\ref{eq:rigid-pair-product} has a known
transformation law, the resulting edge features remain equivariant under
global rotation,
\begin{equation}
    \mathcal Z_{ij}^{(L)}
    \rightarrow
    D^{(L)}(S)
    \mathcal Z_{ij}^{(L)}.
\end{equation}
At the same time, an independent molecular symmetry operation applied to
either body leaves the rigid-pair feature unchanged,
\begin{equation}
    \mathcal Z_{ij}^{\mathrm{rigid}}
    (R_i g_i,R_j g_j)
    =
    \mathcal Z_{ij}^{\mathrm{rigid}}
    (R_i,R_j),
    \qquad
    g_i\in G_i,\;
    g_j\in G_j.
\end{equation}

The resulting rigid-body tensor products are projected into copies of the
same angular-momentum irreducible representations used by the ordinary MACE
edge spherical harmonics. The complete edge attribute supplied to the MACE
interaction stack is then
\begin{equation}
    \mathbf e_{ij}^{\mathrm{ang}}
    =
    Y(\hat{\mathbf r}_{ij})
    \oplus
    \widetilde{\mathcal Z}_{ij}^{\mathrm{rigid}},
    \label{eq:expanded-mace-edge}
\end{equation}
where
$\widetilde{\mathcal Z}_{ij}^{\mathrm{rigid}}$ denotes the projected
rigid-body edge representation.

Thus the modification leaves the central MACE message-passing architecture
unchanged. Rather than replacing the interaction model, the rigid-body
extension enlarges the geometric information available on each edge:
\begin{widetext}
\begin{equation}
    \underbrace{
    Y(\hat{\mathbf r}_{ij})
    }_{\mathrm{standard\ MACE}}
    \quad\longrightarrow\quad
    \underbrace{
    Y(\hat{\mathbf r}_{ij})
    }_{\mathrm{positional\ geometry}}
    \oplus
    \underbrace{
    \widetilde{\mathcal Z}_{ij}^{\mathrm{rigid}}
    }_{\mathrm{relative\ orientational\ geometry}}.
\end{equation}
\end{widetext}
These expanded edge irreducible representations are subsequently used by the
standard MACE interaction blocks to construct higher-order, many-body
equivariant features and ultimately a scalar energy prediction.

\subsection{Datasets}

\subsubsection{Generation of the random Gay-Berne dimer dataset}
\label{sec:random_rotations_gb_generation}

The dataset \texttt{gay\_berne.xyz} was generated as an
extended XYZ trajectory of isolated two-particle Gay-Berne dimers. The displacement vectors between particles were sampled independently from a uniform
distribution over
\[
    r_\alpha \sim \mathcal U(\sigma_0,2\sigma_0),
    \qquad \alpha\in\{x,y,z\},
\]
where
\[
    \sigma_0 = 1.
\]

The two particle orientations were sampled independently and uniformly over
the rotation group using \texttt{scipy.spatial.transform.Rotation.random()}.
Each ellipsoid had principal dimensions
\[
    (a_0,b_0,c_0) = (1,1.5,2),
\]
The characteristic Gay--Berne length was taken to be the smallest principal
dimension,
\[
    \sigma_0 = \min(a_0,b_0,c_0)=1.
\]
The orientation-dependent energy-anisotropy parameters were constructed from
the particle dimensions as
\[
    e_a = \sigma_0 \frac{a_0}{b_0c_0},
    \qquad
    e_b = \sigma_0 \frac{b_0}{a_0c_0},
    \qquad
    e_c = \sigma_0 \frac{c_0}{b_0a_0},
\]
The same shape and energy-anisotropy parameters were used for both particles. Potential energies were computed via LAMMPS\cite{thompson_lammps_2022} and reported in reduced units. Forces and torques were computed via centered finite differences on particle positions and orientations, respectively, with all energy calls conducted by LAMMPs. After energy, forces, and torques were computed, configurations were retained only if $U < 0.1$. This filter removed strongly repulsive configurations while retaining
attractive and weakly repulsive structures. Each computed frame contains the simulation cell, particle positions, scalar energy, forces, torques, quaternions, and the three ellipsoid-diameter arrays.

\subsubsection{Generation and evaluation of formamide, benzene, and water clusters}

Dimer and trimer configurations were generated using a custom Python workflow built with the Atomic Simulation Environment (ASE). A near-planar gas-phase geometry was obtained from the experimental gas-phase structure(s) compiled in the NIST Computational Chemistry Comparison and Benchmark Database, Standard Reference Database 101\cite{johnson_nist_1999}. 

To introduce limited intramolecular flexibility, each molecule was independently perturbed by small random Cartesian displacements. The displacement amplitudes were approximately mass weighted, such that hydrogen atoms underwent larger displacements than the heavier atoms. Additional correlated displacements were applied along bond directions to sample small bond-stretching and bending distortions. The resulting perturbation was scaled to a target root-mean-square atomic displacement of 0.035\AA, with the displacement of any individual atom capped at 0.10\AA. The original molecular center of mass was restored after distortion. These perturbations were intended to provide controlled sampling near the equilibrium geometry and were not treated as an exact thermal normal-mode distribution.

Cluster geometries were generated by independently rotating each distorted monomer using uniformly sampled three-dimensional rotation matrices and placing the molecular centers of mass at randomly selected relative positions. Both compact and weakly interacting configurations were included. Center-of-mass separations were sampled using a mixture distribution that preferentially sampled compact arrangements near typical $\pi$-stacking or hydrogen-bonding distances while retaining a broader tail of more weakly interacting configurations. For trimers, the third molecule was placed relative to either the first or second molecule, thereby producing triangular, chain-like, and partially dissociated arrangements. Configurations containing severe intermolecular atomic overlaps were rejected using minimum allowed interatomic distances. After assembly, each complete cluster was translated so that its total center of mass was located at the origin. All calculations were performed under nonperiodic boundary conditions.

\begin{figure*}[t]
    \centering
    \includegraphics[width=0.75\linewidth]{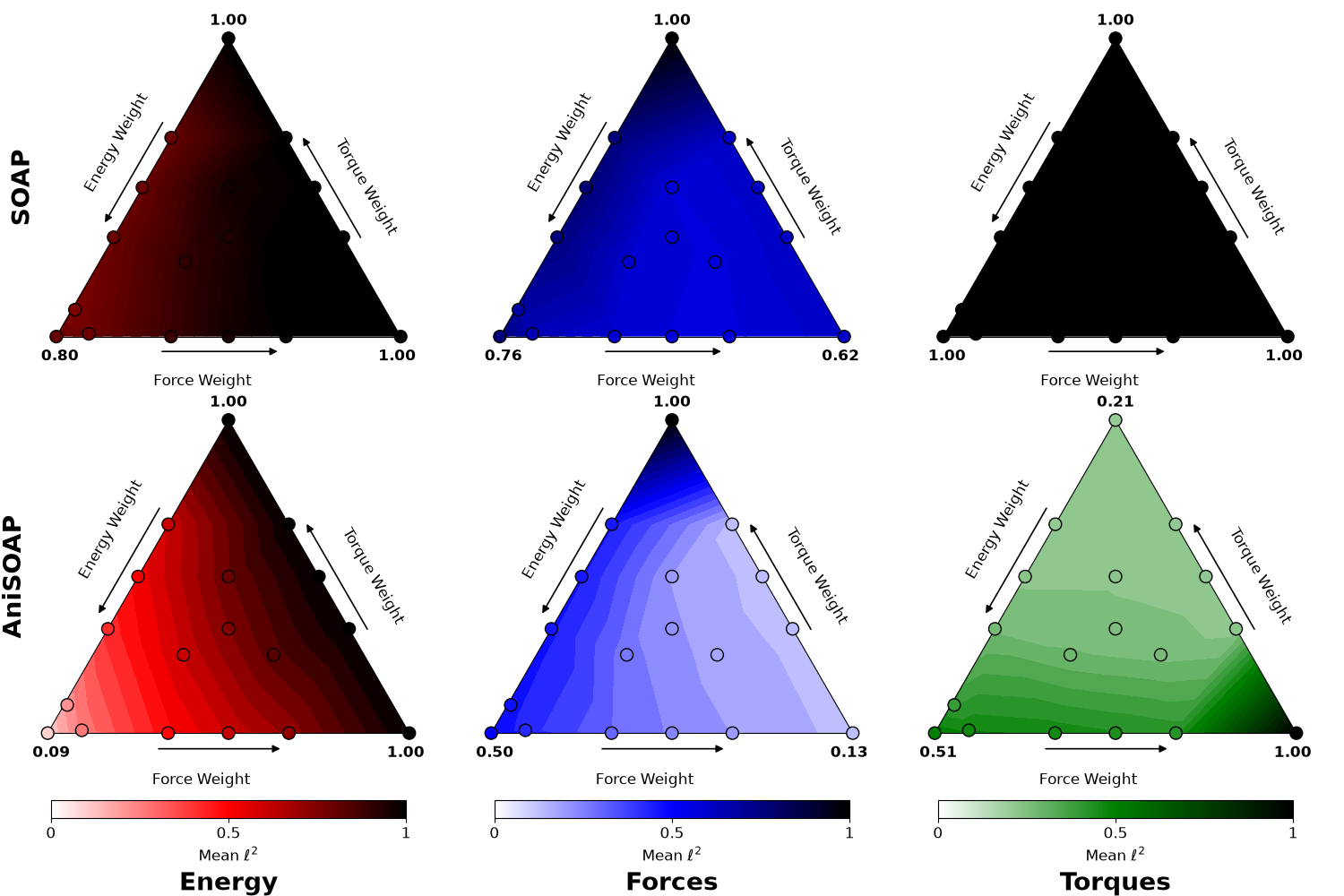}
    \caption{
    Energy, force, and torque tradeoff in performance (through $l^2 \equiv \frac{||y _\text{true} - y_\text{pred}||}{||y_\text{true}||}$) for the linear SOAP and \anisoap~models for the two-particle Gay-Berne dataset. As isotropic models (e.g., SOAP) cannot provide angular derivatives, the torque error will always be $l^2=1.$ Scatter points indicate weights that were explicitly tested, the background colors indicate the interpolated values. Each weight was tested for 5 independent train/test splits; values shown are performance metrics for the test set averaged across all splits.
    }
    \label{fig:linear_anisoap_learning_weight_tradeoff_gb}
\end{figure*}

For each separate chemistry, a total of 10,000 cluster configurations were generated; a fixed pseudorandom seed was used to make the configuration generation reproducible.

Potential energies and atomic forces were evaluated using the open-source FAIR Chemistry implementation of the UMA universal machine-learned interatomic potential\cite{wood_uma_2025}. The `uma-s-1p2' pretrained checkpoint was used with the OMol task through the `FAIRChemCalculator' interface. Each cluster was treated as a neutral singlet with total charge zero and spin multiplicity one. For each evaluated cluster, the total potential energy and Cartesian atomic forces were stored using an ASE `SinglePointCalculator'. Molecular centers of mass were calculated from the instantaneous distorted atomic coordinates and atomic masses. The Cartesian inertia tensor of each molecule was calculated about its instantaneous center of mass $\mathbf{R}_m$ as
\begin{equation} \mathbf{I}_m = \sum_{i \in m} m_i \left[ \left\lVert \mathbf{r}_i-\mathbf{R}_m \right\rVert^2 \mathbf{I} - \left(\mathbf{r}_i-\mathbf{R}_m\right) \left(\mathbf{r}_i-\mathbf{R}_m\right)^{\mathrm{T}} \right]. \label{eq:inertia-tensor} \end{equation}
where $m_i$ and $\mathbf{r}_i$ are the mass and Cartesian position of atom (i), respectively. $\mathbf{I}$ is the \(3 \times 3\) identity tensor. Diagonalization of \(\mathbf{I}_m\) yielded the three principal moments of inertia and the corresponding principal axes.
The net force acting on each molecule was obtained by summing the atomic forces belonging to that molecule,

\begin{equation} \mathbf{F}_m = \sum_{i \in m} \mathbf{F}_i. \label{eq:molecular-force} \end{equation}

The molecular torque was calculated with respect to the instantaneous molecular center of mass,

\begin{equation} \boldsymbol{\tau}_m = \sum_{i \in m} \left(\mathbf{r}_i-\mathbf{R}_m\right) \times \mathbf{F}_i. \label{eq:molecular-torque} \end{equation}
Because the torque was evaluated relative to the molecular center of mass, it was invariant under an arbitrary global translation of the cluster. When energy decomposition was enabled, each distorted monomer was also evaluated in isolation without changing its internal geometry. For a cluster containing \(N\) molecules, the total interaction energy was defined as \begin{equation} E_{\mathrm{int}} = E_{1\ldots N} - \sum_{i=1}^{N} E_i. \label{eq:interaction-energy} \end{equation} Here, \(E_{1\ldots N}\) is the energy of the complete cluster and \(E_i\) is the energy of isolated distorted monomer \(i\). For a dimer composed of molecules \(i\) and \(j\), the pair interaction energy was calculated as \begin{equation} E_{ij}^{(2)} = E_{ij} - E_i - E_j. \label{eq:pair-interaction} \end{equation} For a trimer, the nonadditive three-body contribution was calculated as \begin{equation} E_{123}^{(3)} = E_{123} - E_{12} - E_{13} - E_{23} + E_1 + E_2 + E_3. \label{eq:three-body-energy} \end{equation} Here, \(E_{123}\) is the energy of the complete trimer, \(E_{ij}\) is the energy of the distorted dimer containing molecules \(i\) and \(j\), and \(E_i\) is the energy of isolated distorted monomer \(i\).

\section{Results}

\subsection{Random Gay-Berne Dimers}
\label{sec:results_linear_anisoap_gb}

As a first pass, we evaluated how each model can be used to predict energies, forces, and torques for the random
Gay-Berne dimer data set. All models in this section use the same fixed train/validation/test split. The split contains 1570 training configurations, 316 validation configurations, and 388 held-out test configurations. Feature selection, feature scaling, and target normalization were fit using only the training set. Model parameters were tuned on the validation set, after which each model was refit on the combined training and validation sets and evaluated once on the held-out test set.

\begin{table*}[t]
\caption{
Comparison of two derivative-evaluation procedures for the linear \anisoap~model trained on energies. Both
models use $(\beta_E,\beta_F,\beta_\Tau)=(1,0,0)$. In the finite-difference evaluation,
forces and torques are obtained by finite differences of the learned scalar
energy. In the descriptor-derivative evaluation, forces and torques are
obtained by contracting precomputed descriptor derivatives with the learned
linear coefficients. 
}
\label{tab:linear_anisoap_derivative_evaluation}
\begin{ruledtabular}
\begin{tabular}{lcccccc}
Derivative evaluation &
$R^2_E$ &
$R^2_F$ &
$R^2_\Tau$ &
RMSE$_E$ ($\varepsilon$)&
RMSE$_F$ ($\varepsilon / \sigma_0$)&
RMSE$_\Tau$ ($\varepsilon$)\\
\hline
Finite differences of learned energy &
0.9938 &
0.7479 &
0.7460 &
$9.93\times 10^{-4}$ &
0.03424 &
0.01098 \\

Descriptor derivatives &
0.9913 &
0.7485 &
0.7441 &
$1.18\times 10^{-3}$ &
0.03420 &
0.01103 \\
\end{tabular}
\end{ruledtabular}
\end{table*}

First, we demonstrate the equivalence of computing forces and torques through finite-difference of the learned scalar energy and through direct contraction of descriptor derivatives with the learned linear coefficients (Table~\ref{tab:linear_anisoap_derivative_evaluation}). Both models use $(\beta_E,\beta_F,\beta_\tau)=(1,0,0)$ (i.e. the $\beta$ weights indicate only training on energies) and give nearly identical held-out performance. The finite-difference evaluation achieves $R_E^2=0.9938$, $R_F^2=0.7479$, and $R_\tau^2=0.7460$, while the descriptor-derivative evaluation gives $R_E^2=0.9913$, $R_F^2=0.7485$, and $R_\tau^2=0.7441$. This agreement shows that, for the linear \anisoap~model, forces and torques can be evaluated from descriptor derivatives without loss of accuracy relative to finite-differencing the learned energy. At the same time, both models predict energies more accurately than derivatives, indicating that accurate interpolation of the scalar energy surface alone does not fully learn the force and torque fields.

As expected, the representative isotropic model performs poorly in the learning of energies and forces, with the former optimized at $(\beta_E,\beta_F,\beta_\tau)=(1,0,0)$ with  $l_E^2 \equiv \frac{||E _\text{true} - E_\text{pred}||}{||E_\text{true}||}=0.8$, and the latter at $(\beta_E,\beta_F,\beta_\tau)=(0,1,0)$ with $l_F^2=0.62$. Isotropic models, fundamentally, cannot provide angular derivatives, and thus the torque error will always be $l_\tau^2=1$.
With the anisotropic model (\anisoap), we see improvement in all three metrics, with $min(l_E^2)=0.09$, $min(l_F^2)=0.13$ and $min(l_\Tau^2)=0.21$, all at their respective independent learning exercises.
Varying the learning weights $(\beta_E,\beta_F,\beta_\tau)$ reveals a clear tradeoff between energy accuracy and derivative accuracy (Fig.~\ref{fig:linear_anisoap_learning_weight_tradeoff_gb}). Increasing the relative contribution of force and torque labels improves the mean force/torque test $l^2 \equiv \frac{||y _\text{true} - y_\text{pred}||}{||y_\text{true}||}$. The intermediate weighting regime provides a useful compromise, improving both forces and torques relative to the energy-trained model while maintaining moderate $l_E^2$. The force-trained and torque-trained single-observable models achieve high accuracy for the observable included in the objective, but poor performance on the others, emphasizing that balanced energy, force, and torque prediction requires fitting a shared scalar model with multiple observables. Full error tables are provided in the Supporting Information.

\subsection{Linear \anisoap~baselines on molecular dimers and trimers}
\label{sec:results_linear_anisoap_molecules}

\begin{figure}[t]
    \centering
    \begin{subfigure}{\linewidth}
    \caption{Benzene}
    \includegraphics[width=\linewidth, trim=0 50 0 0, clip]{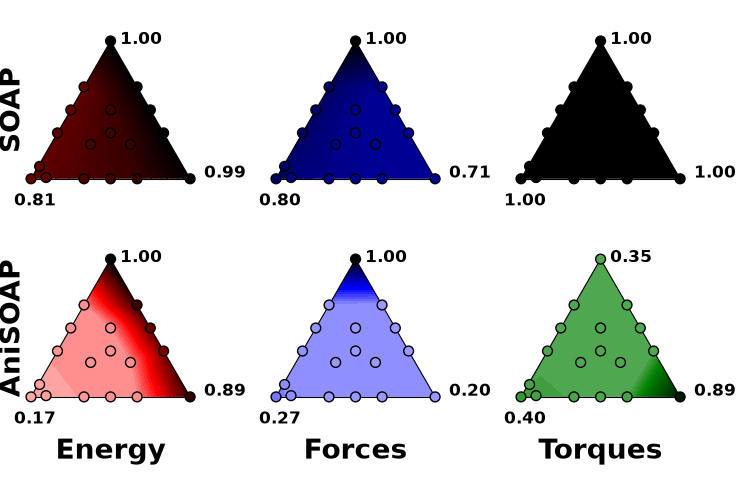}
    \end{subfigure}
    \begin{subfigure}{\linewidth}
    \caption{Water}
    \includegraphics[width=\linewidth, trim=0 50 0 0, clip]{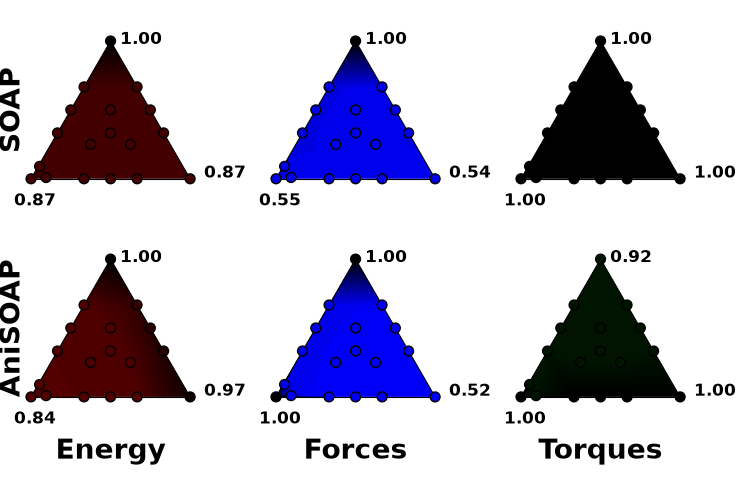}
    \end{subfigure}
    \begin{subfigure}{\linewidth}
    \caption{Formamide}
    \includegraphics[width=\linewidth, trim=0 50 0 0, clip]{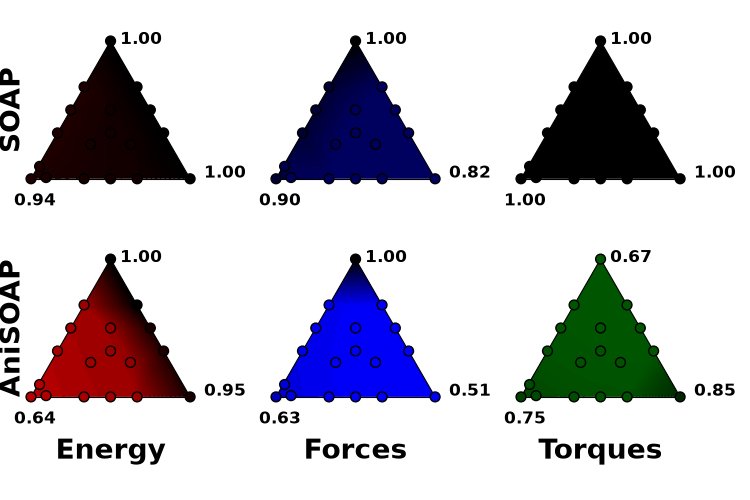}
    \end{subfigure}
    \caption{
    Energy (left), force (center), and torque (right) tradeoff in performance for molecular targets. For each plot, the energy, force, and torque training weights vary in the same way as Fig.~\ref{fig:linear_anisoap_learning_weight_tradeoff_gb}, with the same colorscales. 
    }
    \label{fig:linear_anisoap_learning_weight_tradeoff}
\end{figure}
We next evaluated the same linear \anisoap~framework on a chemically realistic
coarse-grained data sets. In contrast to the Gay--Berne dimers, where
the target energies, forces, and torques are generated from an analytic
anisotropic pair potential, these targets are obtained by first
evaluating atomistic cluster energies and atomic forces and then projecting
those quantities onto molecular centers of mass. 

Our goal in this section is not to identify the most expressive possible potential, but to establish what information is recoverable from the
chosen coarse-grained variables using a simple, interpretable conservative
model. These linear \anisoap~models therefore serve as a representation-level
baseline for the more expressive anisotropic MACE models considered below.
If the linear models fail completely, this would suggest that the retained
coarse-grained variables do not encode the relevant molecular information.
Conversely, nontrivial linear performance indicates that the position,
orientation, and geometry variables contain recoverable energy, force, and
torque signal, even if a more flexible architecture is needed to exploit that
signal fully.

Within each dataset, each molecule was represented as a single anisotropic coarse-grained
particle located at its molecular center of mass. Particle orientations were
defined by the principal axes of the molecular inertia tensor, and ellipsoid
diameters were inferred from the corresponding principal moments. Unless
otherwise stated, the \anisoap~descriptor used $n_{\max}=10$, $l_{\max}=10$,
cutoff radius $r_{\mathrm{cut}}=8.0~\mathrm{\AA}$, radial Gaussian width
$2.5~\mathrm{\AA}$, and a uniform ellipsoid-diameter scale factor of 0.7.
These descriptor parameters define a physically constrained, moderately
smoothed representation of the molecular shape. All models were trained and
evaluated on the same fixed train/validation/test split, with feature
selection, feature scaling, and target normalization fit using only the
training set.

Figure~\ref{fig:linear_anisoap_learning_weight_tradeoff} compares the
energy, force, and torque errors obtained with isotropic SOAP and \anisoap\
across benzene, formamide, and water. As in the Gay--Berne benchmark, the
vertices of each triangle correspond to models trained exclusively on energy,
force, or torque, while interior points correspond to joint fits with mixed
learning weights. The interpolated surfaces therefore illustrate
both the best performance available for each observable and the tradeoff
between fitting energies and their translational and rotational derivatives.

The benefit of explicitly retaining molecular anisotropy is strongest for
benzene. At the single-observable limits, the normalized energy error
decreases from $l_E^2=0.81$ with SOAP to $0.17$ with \anisoap, while the
force error decreases from $l_F^2=0.71$ to $0.19$. More importantly,
isotropic SOAP contains no orientational degree of freedom and consequently
gives $l_\tau^2=1$ throughout the torque panel. \anisoap, in contrast,
recovers substantial torque information, reaching $l_\tau^2=0.33$ when
trained directly on torques and $l_\tau^2=0.38$ even at the energy-trained
vertex. The broad low-error regions in the \anisoap\ energy and force
surfaces further indicate that the anisotropic descriptor retains information
useful to more than one mechanical observable simultaneously.

Formamide shows the same qualitative behavior, although with substantially
larger residual errors. The energy-only error decreases from $0.94$ for SOAP
to $0.64$ for \anisoap, and the force-optimized error decreases from $0.81$
to $0.51$. The anisotropic representation again introduces recoverable
rotational information: whereas SOAP is constrained to
$l_\tau^2=1$, the torque-trained \anisoap\ model reaches
$l_\tau^2=0.67$. Thus, even for the more chemically heterogeneous formamide
clusters, molecular shape and orientation provide information that is absent
from a center-of-mass-only isotropic representation. At the same time, the
higher errors relative to benzene indicate that an ellipsoidal model
does not capture the projected molecular interaction surface equally well
for all chemistries.

Water provides an interesting case. AniSOAP provides only modest improvements, with energy-trained
errors of $0.88$ and $0.83$ for SOAP and \anisoap, respectively. The
force-optimized errors are likewise similar, $0.55$ for SOAP and $0.51$ for
\anisoap. Moreover, the torque information recovered by \anisoap\ is weak:
even direct torque training gives $l_\tau^2=0.91$, only slightly below the
isotropic value of unity. The water results therefore show that introducing
an anisotropic density alone does not guarantee that the chosen coarse-grained
geometry provides a useful representation of molecular orientation.

Taken together, these three systems establish a hierarchy in the usefulness
of the ellipsoidal \anisoap\ representation. Benzene, which is reasonably approximated by ellipsoidal symmetry, shows a large gain from
retaining anisotropy, formamide, with no symmetry, shows a clear but more moderate gain, and
water shows little improvement over the isotropic baseline. These results
reinforce the role of \anisoap\ as a representation-level diagnostic: the
descriptor can expose energy, force, and torque information when that
information is captured by the chosen molecular geometry, but additional
representational or architectural structure is required when it is not.
This observation motivates the equivariant rigid-body MACE construction
considered in the following section.
\subsection{Rigid-body equivariant MACE models}
\label{sec:results_rigid_mace}

We next tested whether explicit rigid-body information could be incorporated
directly into an equivariant message-passing potential, taking the clearest case of failure for \anisoap, water, as our test case. For each molecular
system, we compared five representations: the best linear AniSOAP model, an
isotropic MACE model using only center-of-mass positions and species, MACE
augmented with moment-of-inertia (MOI) features, MACE augmented with
orientation-dependent rigid-pair features, and MACE containing both MOI and
orientation-dependent features. For water, the orientational features use the symmetry-adapted $C_2$ representation described in
Sec.~\ref{sec:rigid_mace}; benzene and formamide use the full molecular frame.

Unless otherwise stated, the MACE models were optimized using energy and force
labels. Molecular torques therefore provide a particularly stringent test of
whether the learned scalar energy has acquired the correct orientational
dependence: a model whose energy is independent of molecular orientation
cannot produce a nonzero conservative torque.

\begin{figure*}[t]
    \centering
    \includegraphics[width=\linewidth, clip, trim=0 0 0 80]{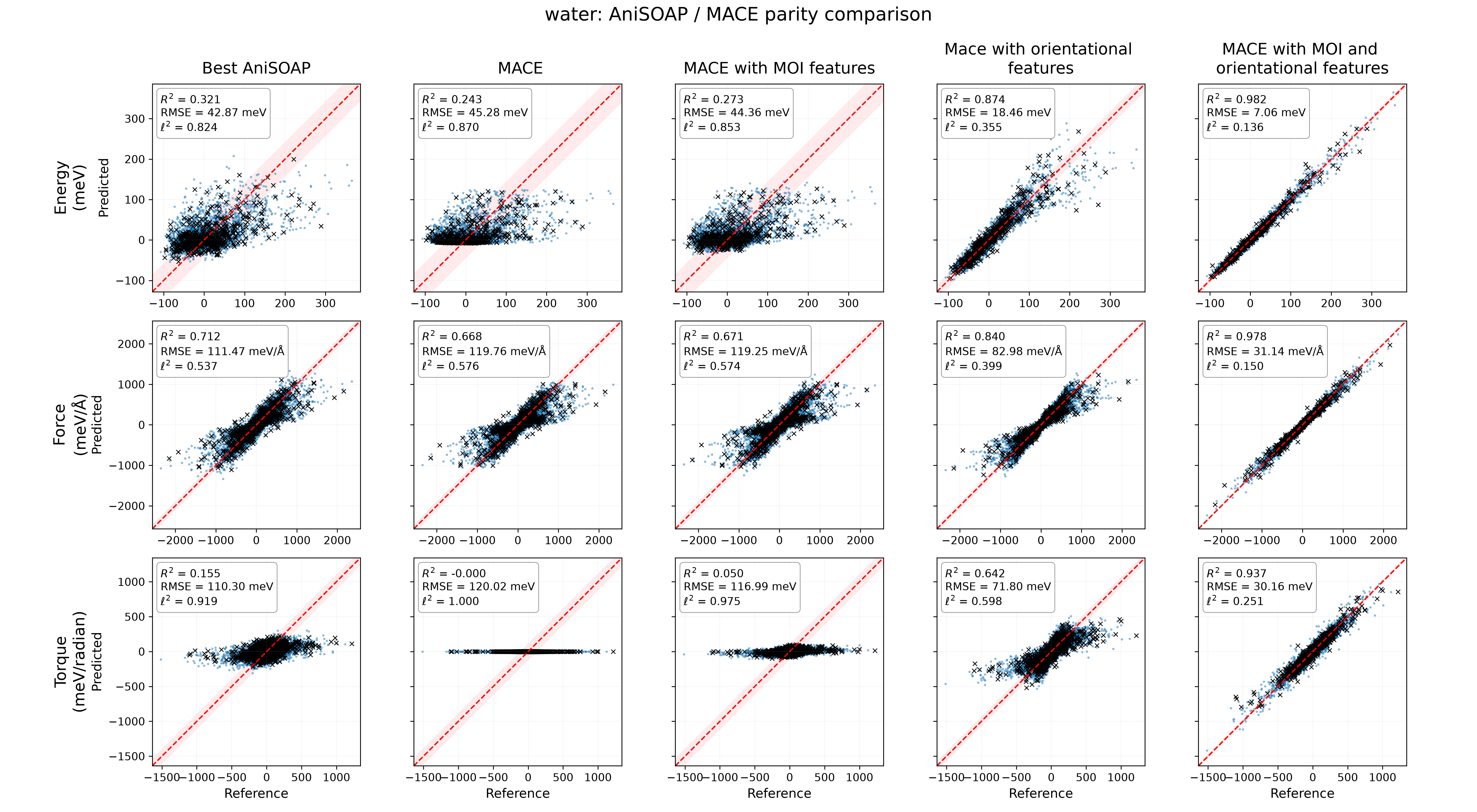}
    \caption{Parity plots of the energy, force, and torque learning tasks for water clusters. Models were trained solely on energy and force data. The left column displays the best AniSOAP results, and the other columns are the MACE models. As MACE incorporates additional features (MOI, $C_2$ symmetry, and MOI+$C_2$ symmetry), the regressions systematically improve. }
    \label{fig:water_comparison}
\end{figure*}
\subsection{Water: molecular orientation resolves a representation bottleneck}

Water exhibits the clearest dependence on the retained orientational degrees
of freedom, as shown in Figure \ref{fig:water_comparison}. The isotropic MACE model performs similarly to the linear SOAP
baseline for energies and forces, but contains essentially no torque signal.
The isotropic MACE model gives test scores of
$R_E^2=0.243$, $R_F^2=0.668$, and $R_\tau^2\approx0$, with corresponding
relative $\ell_2$ errors of 0.870, 0.576, and 1.000. Similar to the difference between SOAP and \anisoap~models, adding the MOI features
alone produces almost no improvement: $R_E^2=0.273$, $R_F^2=0.671$, and
$R_\tau^2=0.050$. Thus, despite encoding a second-rank description of
molecular geometry, the inertia tensor does not provide sufficient orientational information to reproduce the water interaction surface in this benchmark.

The behavior changes qualitatively when explicit rigid-pair orientation is
introduced. The symmetry-adapted $C_2$ pair representation increases the test
scores to $R_E^2=0.874$, $R_F^2=0.840$, and $R_\tau^2=0.642$, reducing the
energy RMSE from 45.28 to 18.46 meV (per bead) and the torque RMSE from 120.02 to 71.80 meV (per bead). The combined MOI and $C_2$-orientation model performs substantially better still, reaching $R_E^2=0.982$, $R_F^2=0.978$, and $R_\tau^2=0.937$. Its RMSE values are 7.06 meV for energy, 31.14 meV/\AA{} for force components, and 30.16 meV for torque components. Relative to isotropic MACE, these correspond to reductions in RMSE of approximately 84\%, 74\%, and 75\% for energies, forces, and torques,
respectively.

The contrast between the MOI-only and orientation-aware models is particularly important. A generic inertia tensor possesses a $D_2$ rotational stabilizer, whereas the proper rotational symmetry of water is only $C_2$. MOI therefore identifies physically inequivalent water orientations and cannot represent the complete rigid-water orientation space. The $C_2$ representation explicitly removes only the physical molecular equivalence while retaining the orientational distinctions required by the interaction energy. The production results are therefore consistent with the controlled water benchmark, in which isotropic and MOI representations were indistinguishable while full-frame and symmetry-adapted orientational features resolved the missing energy variation.

Interestingly, the production model containing both MOI and $C_2$ pair
features performs better than the $C_2$ pair representation alone. This does
not imply that MOI becomes a complete orientation representation. Rather, the low-order tensor features appear to provide a useful complementary geometric signal once the false orientation aliases have already been resolved by the $C_2$ pair representation. Thus, the two feature classes play different roles: the symmetry-adapted frame establishes orientational
completeness, while the MOI channels provide an economical low-order
description that can facilitate learning.

\section{Conclusions}
Machine-learned interatomic potentials have become increasingly powerful and accurate for atomistic systems, but are still limited in accurately describing the interactions between coarse-grained sites, especially if the molecules grouped into sites exhibit anisotropic interactions. In this work, we demonstrated that incorporating molecular anisotropy into the SOAP and MACE formalisms (\anisoap, MACE-CG) can systematically learn the energies, forces, and torques of various single-site anisotropic coarse-grained particles. \anisoap~serves as a simpler baseline, utilizing linear models on the anisotropic features to predict the targets. AniSOAP incorporates $D_v$ (ellipsoidal) molecular symmetry, enabling the accurate learning of ellipsoidal bodies, such as benzene molecules and Gay-Berne particles. However, for molecules with other molecular symmetries, such as water, the AniSOAP formalisms fail to accurately learn the targets. 

The flexibility of the MACE formalism enables the integration of different molecular geometries as node features. As a baseline comparison, we demonstrate that MACE alone cannot accurately reproduce energies, forces, and torques for water molecules, as MACE is designed for isotropic particles. We incorporated moment-of-inertia features into the MACE model, which is analogous to defining the ellipsoidal geometry in \anisoap, and obtain similar performance as the \anisoap~models. However, incorporating water's $C_2$ point-group symmetry into our MACE-CG model dramatically improved the models, indicating the importance of including molecular point-group symmetry when performing single-site coarse-graining.

By its nature, coarse-grained models incur inaccuracy in exchange for simpler, more computationally efficient models. This work has demonstrated that for single-site coarse-graining, we can systematically increase CG models' accuracy by incorporating the molecular geometry. This suggests future directions for coarse-graining, where model accuracy can be tuned in two ways - the usual way of tuning the resolution of the coarse-grained models (how many beads are utilized to represent the molecular system), as well the amount of geometric/symmetry priors to incorporate for each bead.

\bibliography{references}

@misc{johnson_nist_1999,
  doi = {10.18434/T47C7Z},
  url = {http://cccbdb.nist.gov/},
  author = {Johnson,  RD},
  language = {en},
  title = {Computational Chemistry Comparison and Benchmark Database,  NIST Standard Reference Database 101},
  publisher = {National Institute of Standards and Technology},
  year = {2002},
  copyright = {License Information for NIST data}
}

@article{batatia_foundation_2025,
    title = {A foundation model for atomistic materials chemistry},
    volume = {163},
    issn = {0021-9606},
    url = {https://doi.org/10.1063/5.0297006},
    doi = {10.1063/5.0297006},
    number = {18},
    urldate = {2026-06-24},
    journal = {The Journal of Chemical Physics},
    author = {Batatia, Ilyes and Benner, Philipp and Chiang, Yuan and Elena, Alin M. and Kovács, Dávid P. and Riebesell, Janosh and Advincula, Xavier R. and Asta, Mark and Avaylon, Matthew and Baldwin, William J. and Berger, Fabian and Bernstein, Noam and Bhowmik, Arghya and Bigi, Filippo and Blau, Samuel M. and Cărare, Vlad and Ceriotti, Michele and Chong, Sanggyu and Darby, James P. and De, Sandip and Della Pia, Flaviano and Deringer, Volker L. and Elijošius, Rokas and El-Machachi, Zakariya and Fako, Edvin and Falcioni, Fabio and Ferrari, Andrea C. and Gardner, John L. A. and Gawkowski, Mikołaj J. and Genreith-Schriever, Annalena and George, Janine and Goodall, Rhys E. A. and Grandel, Jonas and Grey, Clare P. and Grigorev, Petr and Han, Shuang and Handley, Will and Heenen, Hendrik H. and Hermansson, Kersti and Ho, Cheuk Hin and Hofmann, Stephan and Holm, Christian and Jaafar, Jad and Jakob, Konstantin S. and Jung, Hyunwook and Kapil, Venkat and Kaplan, Aaron D. and Karimitari, Nima and Kermode, James R. and Kourtis, Panagiotis and Kroupa, Namu and Kullgren, Jolla and Kuner, Matthew C. and Kuryla, Domantas and Liepuoniute, Guoda and Lin, Chen and Margraf, Johannes T. and Magdău, Ioan-Bogdan and Michaelides, Angelos and Moore, J. Harry and Naik, Aakash A. and Niblett, Samuel P. and Norwood, Sam Walton and O’Neill, Niamh and Ortner, Christoph and Persson, Kristin A. and Reuter, Karsten and Rosen, Andrew S. and Rosset, Louise A. M. and Schaaf, Lars L. and Schran, Christoph and Shi, Benjamin X. and Sivonxay, Eric and Stenczel, Tamás K. and Sutton, Christopher and Svahn, Viktor and Swinburne, Thomas D. and Tilly, Jules and van der Oord, Cas and Vargas, Santiago and Varga-Umbrich, Eszter and Vegge, Tejs and Vondrák, Martin and Wang, Yangshuai and Witt, William C. and Wolf, Thomas and Zills, Fabian and Csányi, Gábor},
    month = nov,
    year = {2025},
    pages = {184110},
}

@article{noid_rigorous_2024,
    title = {Rigorous {Progress} in {Coarse}-{Graining}},
    volume = {75},
    issn = {0066-426X, 1545-1593},
    url = {https://www.annualreviews.org/content/journals/10.1146/annurev-physchem-062123-010821},
    doi = {10.1146/annurev-physchem-062123-010821},
    language = {en},
    number = {Volume 75, 2024},
    urldate = {2026-06-17},
    journal = {Annual Review of Physical Chemistry},
    publisher = {Annual Reviews},
    author = {Noid, W. G. and Szukalo, Ryan J. and Kidder, Katherine M. and Lesniewski, Maria C.},
    month = jun,
    year = {2024},
    pages = {21--45},
}

@article{gorlich_mapping_2026,
    title = {Mapping {Still} {Matters}: {Coarse}-{Graining} with {Machine} {Learning} {Potentials}},
    volume = {66},
    issn = {1549-9596},
    shorttitle = {Mapping {Still} {Matters}},
    url = {https://doi.org/10.1021/acs.jcim.5c03035},
    doi = {10.1021/acs.jcim.5c03035},
    number = {4},
    urldate = {2026-05-15},
    journal = {Journal of Chemical Information and Modeling},
    publisher = {American Chemical Society},
    author = {Görlich, Franz and Zavadlav, Julija},
    month = feb,
    year = {2026},
    pages = {2166--2176},
}

@article{pozdnyakov_incompleteness_2020,
    title = {Incompleteness of {Atomic} {Structure} {Representations}},
    volume = {125},
    doi = {10.1103/PhysRevLett.125.166001},
    number = {16},
    journal = {Physical Review Letters},
    author = {Pozdnyakov, Sergey N. and Willatt, Michael J. and Bartók, Albert P. and Ortner, Christoph and Csányi, Gábor and Ceriotti, Michele},
    year = {2020},
}

@inproceedings{zhang_aniso-chimes_2026,
	title = {Aniso-{Chimes}: {A} {Generalized} {Machine}-{Learned} {Interaction} {Model} for {Anisotropic} {Nanoparticles}},
	shorttitle = {Aniso-{Chimes}},
	url = {https://aiche.confex.com/aiche/2026/meetingapp.cgi/Paper/734219},
	urldate = {2026-07-14},
	publisher = {AIChE},
	author = {Zhang, Melody Yiyuan and Lee, Shih Kuang and Glotzer, Sharon C. and Lindsey, Rebecca},
	month = nov,
	year = {2026},
}

@article{fakhraei_approximation_2025,
	title = {Approximation of {Anisotropic} {Pair} {Potentials} {Using} {Multivariate} {Interpolation}},
	volume = {129},
	issn = {1520-6106},
	url = {https://doi.org/10.1021/acs.jpcb.5c01451},
	doi = {10.1021/acs.jpcb.5c01451},
	number = {27},
	urldate = {2026-07-14},
	journal = {The Journal of Physical Chemistry B},
	publisher = {American Chemical Society},
	author = {Fakhraei, Mohammadreza and Kieslich, Chris A. and Howard, Michael P.},
	month = jul,
	year = {2025},
	pages = {6985--6996},
}

@article{fakhraei_approximation_2026,
	title = {Approximation of forces and torques from anisotropic pairwise interactions using multivariate polynomials},
	volume = {164},
	issn = {0021-9606},
	url = {https://doi.org/10.1063/5.0318270},
	doi = {10.1063/5.0318270},
	number = {14},
	urldate = {2026-07-14},
	journal = {The Journal of Chemical Physics},
	author = {Fakhraei, Mohammadreza and Bush, Michaela and Kieslich, Chris A. and Howard, Michael P.},
	month = apr,
	year = {2026},
	pages = {144117},
}

@article{wang_machine_2019,
	title = {Machine {Learning} of {Coarse}-{Grained} {Molecular} {Dynamics} {Force} {Fields}},
	volume = {5},
	copyright = {http://pubs.acs.org/page/policy/authorchoice\_termsofuse.html},
	issn = {2374-7943, 2374-7951},
	url = {https://pubs.acs.org/doi/10.1021/acscentsci.8b00913},
	doi = {10.1021/acscentsci.8b00913},
	number = {5},
	urldate = {2025-06-27},
	journal = {ACS Central Science},
	author = {Wang, Jiang and Olsson, Simon and Wehmeyer, Christoph and Perez, Adria and Charron, Nicholas E. and De Fabritiis, Gianni and Noe, Frank and Clementi, Cecilia},
	month = may,
	year = {2019},
	pages = {755--767},
}

@article{marrink_martini_2007,
	title = {The {MARTINI} {Force} {Field}: {Coarse} {Grained} {Model} for {Biomolecular} {Simulations}},
	volume = {111},
	issn = {1520-6106, 1520-5207},
	shorttitle = {The {MARTINI} {Force} {Field}},
	url = {https://pubs.acs.org/doi/10.1021/jp071097f},
	doi = {10.1021/jp071097f},
	number = {27},
	urldate = {2025-06-11},
	journal = {The Journal of Physical Chemistry B},
	author = {Marrink, Siewert J. and Risselada, H. Jelger and Yefimov, Serge and Tieleman, D. Peter and De Vries, Alex H.},
	month = jul,
	year = {2007},
	pages = {7812--7824},
}

@article{behler_neural_2011,
	title = {Neural network potential-energy surfaces in chemistry: {A} tool for large-scale simulations},
	volume = {13},
	doi = {10.1039/c1cp21668f},
	number = {40},
	journal = {Physical Chemistry Chemical Physics},
	author = {Behler, Joerg},
	month = oct,
	year = {2011},
	pages = {17930--17955},
}

@article{unke_machine_2021,
	title = {Machine {Learning} {Force} {Fields}},
	volume = {121},
	copyright = {https://creativecommons.org/licenses/by-nc-nd/4.0/},
	issn = {0009-2665, 1520-6890},
	url = {https://pubs.acs.org/doi/10.1021/acs.chemrev.0c01111},
	doi = {10.1021/acs.chemrev.0c01111},
	number = {16},
	urldate = {2025-02-02},
	journal = {Chem. Rev.},
	author = {Unke, Oliver T. and Chmiela, Stefan and Sauceda, Huziel E. and Gastegger, Michael and Poltavsky, Igor and Schutt, Kristof T. and Tkatchenko, Alexandre and Mueller, Klaus-Robert},
	month = aug,
	year = {2021},
	pages = {10142--10186},
}

@article{hosseini_martini_2024,
	title = {Martini on the {Rocks}: {Can} a {Coarse}-{Grained} {Force} {Field} {Model} {Crystals}?},
	volume = {15},
	shorttitle = {Martini on the {Rocks}},
	url = {https://doi.org/10.1021/acs.jpclett.4c00012},
	doi = {10.1021/acs.jpclett.4c00012},
	number = {4},
	urldate = {2025-06-16},
	journal = {The Journal of Physical Chemistry Letters},
	publisher = {American Chemical Society},
	author = {Hosseini, A. Najla and van der Spoel, David},
	month = feb,
	year = {2024},
	pages = {1079--1088},
}

@article{bartok_representing_2013,
	title = {On representing chemical environments},
	volume = {87},
	copyright = {http://link.aps.org/licenses/aps-default-license},
	issn = {1098-0121, 1550-235X},
	url = {https://link.aps.org/doi/10.1103/PhysRevB.87.184115},
	doi = {10.1103/PhysRevB.87.184115},
	number = {18},
	urldate = {2025-01-29},
	journal = {Phys. Rev. B},
	author = {Bartok, Albert P. and Kondor, Risi and Csanyi, Gabor},
	month = may,
	year = {2013},
	pages = {184115},
}

@misc{wood_uma_2025,
	title = {{UMA}: {A} {Family} of {Universal} {Models} for {Atoms}},
	shorttitle = {{UMA}},
	url = {http://arxiv.org/abs/2506.23971},
	doi = {10.48550/arXiv.2506.23971},
	urldate = {2025-07-11},
	publisher = {arXiv},
	author = {Wood, Brandon M. and Dzamba, Misko and Fu, Xiang and Gao, Meng and Shuaibi, Muhammed and Barroso-Luque, Luis and Abdelmaqsoud, Kareem and Gharakhanyan, Vahe and Kitchin, John R. and Levine, Daniel S. and Michel, Kyle and Sriram, Anuroop and Cohen, Taco and Das, Abhishek and Rizvi, Ammar and Sahoo, Sushree Jagriti and Ulissi, Zachary W. and Zitnick, C. Lawrence},
	month = jun,
	year = {2025},
	note = {arXiv:2506.23971 [cs]},
}

@article{lin_anisoap_2025,
	title = {{AniSOAP}: {Machine} {Learning} {Representations} for {Coarse}-grained and {Non}-spherical {Systems}},
	volume = {10},
	copyright = {http://creativecommons.org/licenses/by/4.0/},
	issn = {2475-9066},
	shorttitle = {{AniSOAP}},
	url = {https://joss.theoj.org/papers/10.21105/joss.07954},
	doi = {10.21105/joss.07954},
	number = {111},
	urldate = {2025-07-11},
	journal = {Journal of Open Source Software},
	author = {Lin, Arthur Yan and Ortengren, Lucas and Hwang, Seonwoo and Cho, Yong-Cheol and Nigam, Jigyasa and Cersonsky, Rose K.},
	month = jul,
	year = {2025},
	pages = {7954},
}

@article{zadok_coarse-grained_2018,
	title = {Coarse-{Grained} {Simulation} of {Protein}-{Imprinted} {Hydrogels}},
	volume = {122},
	issn = {1520-6106},
	url = {https://doi.org/10.1021/acs.jpcb.8b03774},
	doi = {10.1021/acs.jpcb.8b03774},
	number = {28},
	urldate = {2025-06-16},
	journal = {The Journal of Physical Chemistry B},
	publisher = {American Chemical Society},
	author = {Zadok, Israel and Srebnik, Simcha},
	month = jul,
	year = {2018},
	pages = {7091--7101},
}

@article{loose_changing_2024,
	title = {Changing {Your} {Martini} {Can} {Still} {Give} {You} a {Hangover}},
	volume = {20},
	issn = {1549-9618},
	url = {https://doi.org/10.1021/acs.jctc.4c00868},
	doi = {10.1021/acs.jctc.4c00868},
	number = {20},
	urldate = {2025-06-16},
	journal = {Journal of Chemical Theory and Computation},
	publisher = {American Chemical Society},
	author = {Loose, Timothy D. and Sahrmann, Patrick G. and Qu, Thomas S. and Voth, Gregory A.},
	month = oct,
	year = {2024},
	pages = {9190--9208},
}

@article{wilson_anisotropic_2023,
	title = {Anisotropic molecular coarse-graining by force and torque matching with neural networks},
	volume = {159},
	url = {https://pubs.aip.org/aip/jcp/article/159/2/024110/2901764},
	number = {2},
	urldate = {2025-06-10},
	journal = {The Journal of Chemical Physics},
	publisher = {AIP Publishing},
	author = {Wilson, Marltan O. and Huang, David M.},
	year = {2023},
}

@article{anstine_machine_2023,
	title = {Machine {Learning} {Interatomic} {Potentials} and {Long}-{Range} {Physics}},
	volume = {127},
	issn = {1089-5639},
	url = {https://doi.org/10.1021/acs.jpca.2c06778},
	doi = {10.1021/acs.jpca.2c06778},
	number = {11},
	urldate = {2025-06-09},
	journal = {The Journal of Physical Chemistry A},
	publisher = {American Chemical Society},
	author = {Anstine, Dylan M. and Isayev, Olexandr},
	month = mar,
	year = {2023},
	pages = {2417--2431},
}

@article{nguyen_systematic_2022,
	title = {Systematic bottom-up molecular coarse-graining via force and torque matching using anisotropic particles},
	volume = {156},
	issn = {0021-9606},
	url = {https://doi.org/10.1063/5.0085006},
	doi = {10.1063/5.0085006},
	number = {18},
	urldate = {2025-05-30},
	journal = {The Journal of Chemical Physics},
	author = {Nguyen, Huong T. L. and Huang, David M.},
	month = may,
	year = {2022},
	pages = {184118},
}

@article{lin_expanding_2024,
	title = {Expanding density-correlation machine learning representations for anisotropic coarse-grained particles},
	volume = {161},
	issn = {0021-9606},
	url = {https://doi.org/10.1063/5.0210910},
	doi = {10.1063/5.0210910},
	number = {7},
	urldate = {2024-08-21},
	journal = {J. Chem. Phys.},
	author = {Lin, Arthur and Huguenin-Dumittan, Kevin K. and Cho, Yong-Cheol and Nigam, Jigyasa and Cersonsky, Rose K.},
	month = aug,
	year = {2024},
	pages = {074112},
}

@article{pozdnyakov_smooth_2023,
	title = {Smooth, exact rotational symmetrization for deep learning on point clouds},
	volume = {36},
	url = {https://proceedings.neurips.cc/paper_files/paper/2023/hash/fb4a7e3522363907b26a86cc5be627ac-Abstract-Conference.html},
	urldate = {2025-05-29},
	journal = {Advances in Neural Information Processing Systems},
	author = {Pozdnyakov, Sergey and Ceriotti, Michele},
	year = {2023},
	pages = {79469--79501},
}

@article{pozdnyakov_incompleteness_2022,
	title = {Incompleteness of graph neural networks for points clouds in three dimensions},
	volume = {3},
	url = {https://iopscience.iop.org/article/10.1088/2632-2153/aca1f8/meta},
	number = {4},
	urldate = {2025-05-29},
	journal = {Machine Learning: Science and Technology},
	publisher = {IOP Publishing},
	author = {Pozdnyakov, Sergey N. and Ceriotti, Michele},
	year = {2022},
	pages = {045020},
}

@article{noid_multiscale_2008,
	title = {The multiscale coarse-graining method. {I}. {A} rigorous bridge between atomistic and coarse-grained models},
	volume = {128},
	issn = {0021-9606},
	url = {https://doi.org/10.1063/1.2938860},
	doi = {10.1063/1.2938860},
	number = {24},
	urldate = {2025-04-11},
	journal = {The Journal of Chemical Physics},
	author = {Noid, W. G. and Chu, Jhih-Wei and Ayton, Gary S. and Krishna, Vinod and Izvekov, Sergei and Voth, Gregory A. and Das, Avisek and Andersen, Hans C.},
	month = jun,
	year = {2008},
	pages = {244114},
}

@article{thompson_lammps_2022,
	title = {{LAMMPS} - a flexible simulation tool for particle-based materials modeling at the atomic, meso, and continuum scales},
	volume = {271},
	issn = {00104655},
	url = {https://linkinghub.elsevier.com/retrieve/pii/S0010465521002836},
	doi = {10.1016/j.cpc.2021.108171},
	urldate = {2025-02-11},
	journal = {Comp. Phys. Comm.},
	author = {Thompson, Aidan P. and Aktulga, H. Metin and Berger, Richard and Bolintineanu, Dan S. and Brown, W. Michael and Crozier, Paul S. and In 'T Veld, Pieter J. and Kohlmeyer, Axel and Moore, Stan G. and Nguyen, Trung Dac and Shan, Ray and Stevens, Mark J. and Tranchida, Julien and Trott, Christian and Plimpton, Steven J.},
	month = feb,
	year = {2022},
	pages = {108171},
}

@article{behler_generalized_2007,
	title = {Generalized {Neural}-{Network} {Representation} of {High}-{Dimensional} {Potential}-{Energy} {Surfaces}},
	volume = {98},
	doi = {10.1103/PhysRevLett.98.146401},
	number = {14},
	journal = {Phys. Rev. Lett.},
	author = {Behler, Jörg and Parrinello, Michele},
	month = apr,
	year = {2007},
	pages = {146401--146401},
}

@article{drautz_atomic_2019,
	title = {Atomic cluster expansion for accurate and transferable interatomic potentials},
	volume = {99},
	issn = {2469-9950, 2469-9969},
	url = {https://link.aps.org/doi/10.1103/PhysRevB.99.014104},
	doi = {10.1103/PhysRevB.99.014104},
	number = {1},
	urldate = {2025-02-02},
	journal = {Phys. Rev. B},
	author = {Drautz, Ralf},
	month = jan,
	year = {2019},
	pages = {014104},
}

@article{batatia_mace_2022,
	title = {{MACE}: {Higher} {Order} {Equivariant} {Message} {Passing} {Neural} {Networks} for {Fast} and {Accurate} {Force} {Fields}},
	volume = {35},
	shorttitle = {{MACE}},
	url = {https://proceedings.neurips.cc/paper_files/paper/2022/hash/4a36c3c51af11ed9f34615b81edb5bbc-Abstract-Conference.html},
	urldate = {2025-02-11},
	journal = {NeurIPs},
	author = {Batatia, Ilyes and Kovacs, David P. and Simm, Gregor and Ortner, Christoph and Csanyi, Gabor},
	month = dec,
	year = {2022},
	pages = {11423--11436},
}
\appendix
\onecolumngrid

\renewcommand{\thetable}{A\arabic{table}}
\setcounter{table}{0}
\renewcommand{\thefigure}{A\arabic{figure}}
\setcounter{figure}{0}
\begin{table}[p]
\centering
\setlength{\tabcolsep}{5pt}
\renewcommand{\arraystretch}{1.12}
\caption{SOAP and \anisoap~descriptor hyperparameters used for all datasets. The $\alpha$ grid gives the minimum and maximum ridge parameters followed by the number of grid points.}
\label{tab:si_anisoap_hyperparameters}
\begin{tabular}{llcccccccc}
\toprule
Dataset & Descriptor & $n_\mathrm{max}$ & $\ell_\mathrm{max}$ & $r_c$ & $\sigma_r$ & $s_d$ & $r_\mathrm{cond}$ & $t_\mathrm{basis}$ & $\alpha$ grid \\
\midrule
\multirow[c]{2}{*}{\makebox[0pt][c]{\rotatebox[origin=c]{90}{\makebox[0pt][c]{Benzene}}}} & \anisoap & 10 & 10 & 8 & 2.5 & 0.7 & \ensuremath{1\times 10^{-6}} & 0.01 & $[\ensuremath{1\times 10^{-8}},\,\ensuremath{1\times 10^{3}}]$ (24) \\
 & SOAP & 10 & 10 & 8 & 2.5 & 0.7 & \ensuremath{1\times 10^{-6}} & 0.01 & $[\ensuremath{1\times 10^{-8}},\,\ensuremath{1\times 10^{3}}]$ (24) \\
\addlinespace[2pt]
\cmidrule(lr){2-10}
\addlinespace[2pt]
\multirow[c]{2}{*}{\makebox[0pt][c]{\rotatebox[origin=c]{90}{\makebox[0pt][c]{Formamide}}}} & \anisoap & 10 & 10 & 8 & 2.5 & 0.7 & \ensuremath{1\times 10^{-6}} & 0.01 & $[\ensuremath{1\times 10^{-8}},\,\ensuremath{1\times 10^{3}}]$ (24) \\
 & SOAP & 10 & 10 & 8 & 2.5 & 0.7 & \ensuremath{1\times 10^{-6}} & 0.01 & $[\ensuremath{1\times 10^{-8}},\,\ensuremath{1\times 10^{3}}]$ (24) \\
\addlinespace[2pt]
\cmidrule(lr){2-10}
\addlinespace[2pt]
\multirow[c]{2}{*}{\makebox[0pt][c]{\rotatebox[origin=c]{90}{\makebox[0pt][c]{Gay--Berne}}}} & \anisoap & 10 & 10 & 5 & 2 & 1 & \ensuremath{1\times 10^{-6}} & 0.01 & $[\ensuremath{1\times 10^{-12}},\,0.01]$ (41) \\
 & SOAP & 10 & 10 & 5 & 2 & 1 & \ensuremath{1\times 10^{-6}} & 0.01 & $[\ensuremath{1\times 10^{-12}},\,0.01]$ (41) \\
\addlinespace[2pt]
\cmidrule(lr){2-10}
\addlinespace[2pt]
\multirow[c]{2}{*}{\makebox[0pt][c]{\rotatebox[origin=c]{90}{\makebox[0pt][c]{Water}}}} & \anisoap & 10 & 10 & 8 & 2.5 & 0.7 & \ensuremath{1\times 10^{-6}} & 0.01 & $[\ensuremath{1\times 10^{-8}},\,\ensuremath{1\times 10^{3}}]$ (24) \\
 & SOAP & 10 & 10 & 8 & 2.5 & 0.7 & \ensuremath{1\times 10^{-6}} & 0.01 & $[\ensuremath{1\times 10^{-8}},\,\ensuremath{1\times 10^{3}}]$ (24) \\
\bottomrule
\end{tabular}
\end{table}

\begin{table}[p]
\centering
\footnotesize
\setlength{\tabcolsep}{3pt}
\renewcommand{\arraystretch}{1.12}
\caption{Errors for linear SOAP and \anisoap~models for Gay--Berne on the test set. Loss weights are reported as $(w_E,w_F,w_\tau)$. Each error entry is $(\ell^2,\mathrm{RMSE},R^2)$. Torque errors are reported for \anisoap and omitted for isotropic SOAP. Values are mean $\pm$ standard deviation over equivalent runs. Values less than $\ensuremath{0.001}$ are denoted by $\epsilon$ and greater than $\ensuremath{100.0}$ are denoted by $\Omega$.}
\label{tab:si_anisoap_errors_two_particle_gb}
\begin{tabular}{lcccc}
\toprule
 & $(w_E,w_F,w_\tau)$ & $E:\;(\ell^2,\mathrm{RMSE},R^2)$ & $F:\;(\ell^2,\mathrm{RMSE},R^2)$ & $\tau:\;(\ell^2,\mathrm{RMSE},R^2)$ \\
\midrule
\multirow[c]{18}{*}{\makebox[0pt][c]{\rotatebox[origin=c]{90}{\makebox[0pt][c]{\anisoap}}}} & $(0,\,0,\,1)$ & $(\Omega,\,\Omega,\,\ll 0)$ & $(\Omega,\,\Omega,\,\ll 0)$ & $(0.21 \pm \epsilon,\,0.0045 \pm \epsilon,\,0.957 \pm \epsilon)$ \\
 & $(0,\,1,\,0)$ & $(4.7 \pm 5.2,\,0.15 \pm 0.013,\,\ll 0)$ & $(0.13 \pm 0.0011,\,0.0089 \pm \epsilon,\,0.983 \pm \epsilon)$ & $(5.6 \pm 6.5,\,0.21 \pm 0.019,\,\ll 0)$ \\
 & $(0,\,\frac{1}{2},\,\frac{1}{2})$ & $(1,\,0.013,\,-0.105)$ & $(0.13,\,0.0089,\,0.983)$ & $(0.22,\,0.0047,\,0.953)$ \\
 & $(0,\,\frac{1}{3},\,\frac{2}{3})$ & $(1,\,0.013,\,-0.109)$ & $(0.13,\,0.0089,\,0.983)$ & $(0.21,\,0.0047,\,0.954)$ \\
 & $(0,\,\frac{2}{3},\,\frac{1}{3})$ & $(1,\,0.013,\,-0.108)$ & $(0.13,\,0.0091,\,0.982)$ & $(0.22,\,0.0049,\,0.95)$ \\
 & $(1,\,0,\,0)$ & $(0.078 \pm 0.023,\,0.0012 \pm \epsilon,\,0.991 \pm \epsilon)$ & $(0.5 \pm \epsilon,\,0.034 \pm \epsilon,\,0.749 \pm \epsilon)$ & $(0.51 \pm \epsilon,\,0.011 \pm \epsilon,\,0.744 \pm \epsilon)$ \\
 & $(\frac{1}{2},\,0,\,\frac{1}{2})$ & $(0.52,\,0.0066,\,0.73)$ & $(0.45,\,0.031,\,0.794)$ & $(0.23,\,0.005,\,0.947)$ \\
 & $(\frac{1}{2},\,\frac{1}{2},\,0)$ & $(0.61,\,0.0077,\,0.632)$ & $(0.24,\,0.017,\,0.941)$ & $(0.44,\,0.0097,\,0.803)$ \\
 & $(\frac{1}{2},\,\frac{1}{4},\,\frac{1}{4})$ & $(0.6,\,0.0077,\,0.634)$ & $(0.27,\,0.018,\,0.929)$ & $(0.28,\,0.0062,\,0.919)$ \\
 & $(\frac{1}{3},\,0,\,\frac{2}{3})$ & $(0.6,\,0.0076,\,0.635)$ & $(0.46,\,0.031,\,0.793)$ & $(0.21,\,0.0046,\,0.955)$ \\
 & $(\frac{1}{3},\,\frac{1}{3},\,\frac{1}{3})$ & $(0.75,\,0.0094,\,0.443)$ & $(0.21,\,0.014,\,0.955)$ & $(0.26,\,0.0056,\,0.935)$ \\
 & $(\frac{1}{3},\,\frac{2}{3},\,0)$ & $(0.72,\,0.0091,\,0.487)$ & $(0.2,\,0.013,\,0.961)$ & $(0.42,\,0.0092,\,0.821)$ \\
 & $(\frac{1}{4},\,\frac{1}{2},\,\frac{1}{4})$ & $(0.83,\,0.01,\,0.314)$ & $(0.18,\,0.012,\,0.969)$ & $(0.26,\,0.0057,\,0.931)$ \\
 & $(\frac{1}{4},\,\frac{1}{4},\,\frac{1}{2})$ & $(0.79,\,0.01,\,0.372)$ & $(0.2,\,0.013,\,0.962)$ & $(0.23,\,0.005,\,0.947)$ \\
 & $(\frac{2}{3},\,0,\,\frac{1}{3})$ & $(0.42,\,0.0053,\,0.825)$ & $(0.46,\,0.031,\,0.788)$ & $(0.27,\,0.0059,\,0.927)$ \\
 & $(\frac{2}{3},\,\frac{1}{3},\,0)$ & $(0.49,\,0.0061,\,0.765)$ & $(0.3,\,0.02,\,0.91)$ & $(0.46,\,0.01,\,0.785)$ \\
 & $(\frac{9}{10},\,0.009,\,0.09)$ & $(0.21,\,0.0027,\,0.954)$ & $(0.47,\,0.032,\,0.779)$ & $(0.39,\,0.0086,\,0.846)$ \\
 & $(\frac{9}{10},\,0.09,\,0.009)$ & $(0.26,\,0.0033,\,0.93)$ & $(0.42,\,0.029,\,0.822)$ & $(0.47,\,0.01,\,0.778)$ \\
\addlinespace[2pt]
\cmidrule(lr){2-5}
\addlinespace[2pt]
\multirow[c]{18}{*}{\makebox[0pt][c]{\rotatebox[origin=c]{90}{\makebox[0pt][c]{SOAP}}}} & $(0,\,0,\,1)$ & $(0.83 \pm 0.24,\,0.013 \pm \epsilon,\,-0.000207 \pm \epsilon)$ & $(1 \pm \epsilon,\,0.068 \pm \epsilon,\,0 \pm \epsilon)$ &  \\
 & $(0,\,1,\,0)$ & $(0.89 \pm 0.16,\,0.015 \pm \epsilon,\,-0.388 \pm \epsilon)$ & $(0.62 \pm 0.0011,\,0.042 \pm \epsilon,\,0.621 \pm 0.0013)$ &  \\
 & $(0,\,\frac{1}{2},\,\frac{1}{2})$ & $(1,\,0.015,\,-0.387)$ & $(0.61,\,0.042,\,0.624)$ &  \\
 & $(0,\,\frac{1}{3},\,\frac{2}{3})$ & $(1,\,0.015,\,-0.387)$ & $(0.61,\,0.042,\,0.624)$ &  \\
 & $(0,\,\frac{2}{3},\,\frac{1}{3})$ & $(1,\,0.015,\,-0.389)$ & $(0.62,\,0.042,\,0.621)$ &  \\
 & $(1,\,0,\,0)$ & $(0.66 \pm 0.19,\,0.01 \pm \epsilon,\,0.365 \pm 0.0025)$ & $(0.76 \pm 0.0042,\,0.052 \pm \epsilon,\,0.417 \pm 0.0064)$ &  \\
 & $(\frac{1}{2},\,0,\,\frac{1}{2})$ & $(0.8,\,0.01,\,0.362)$ & $(0.77,\,0.052,\,0.409)$ &  \\
 & $(\frac{1}{2},\,\frac{1}{2},\,0)$ & $(0.96,\,0.012,\,0.0812)$ & $(0.57,\,0.039,\,0.671)$ &  \\
 & $(\frac{1}{2},\,\frac{1}{4},\,\frac{1}{4})$ & $(0.9,\,0.011,\,0.19)$ & $(0.6,\,0.041,\,0.644)$ &  \\
 & $(\frac{1}{3},\,0,\,\frac{2}{3})$ & $(0.79,\,0.01,\,0.369)$ & $(0.76,\,0.052,\,0.428)$ &  \\
 & $(\frac{1}{3},\,\frac{1}{3},\,\frac{1}{3})$ & $(0.94,\,0.012,\,0.116)$ & $(0.58,\,0.04,\,0.662)$ &  \\
 & $(\frac{1}{3},\,\frac{2}{3},\,0)$ & $(1,\,0.013,\,-0.0793)$ & $(0.57,\,0.039,\,0.677)$ &  \\
 & $(\frac{1}{4},\,\frac{1}{2},\,\frac{1}{4})$ & $(1,\,0.013,\,-0.0789)$ & $(0.57,\,0.039,\,0.678)$ &  \\
 & $(\frac{1}{4},\,\frac{1}{4},\,\frac{1}{2})$ & $(0.96,\,0.012,\,0.0805)$ & $(0.57,\,0.039,\,0.671)$ &  \\
 & $(\frac{2}{3},\,0,\,\frac{1}{3})$ & $(0.79,\,0.01,\,0.37)$ & $(0.75,\,0.051,\,0.431)$ &  \\
 & $(\frac{2}{3},\,\frac{1}{3},\,0)$ & $(0.89,\,0.011,\,0.215)$ & $(0.59,\,0.04,\,0.651)$ &  \\
 & $(\frac{9}{10},\,0.009,\,0.09)$ & $(0.76,\,0.0096,\,0.426)$ & $(0.68,\,0.046,\,0.537)$ &  \\
 & $(\frac{9}{10},\,0.09,\,0.009)$ & $(0.78,\,0.0099,\,0.388)$ & $(0.64,\,0.044,\,0.586)$ &  \\
\bottomrule
\end{tabular}
\end{table}

\begin{table}[p]
\centering
\footnotesize
\setlength{\tabcolsep}{3pt}
\renewcommand{\arraystretch}{1.12}
\caption{Errors for linear SOAP and \anisoap~models for Benzene on the test set. Loss weights are reported as $(w_E,w_F,w_\tau)$. Each error entry is $(\ell^2,\mathrm{RMSE},R^2)$. Torque errors are reported for \anisoap and omitted for isotropic SOAP. Values are mean $\pm$ standard deviation over equivalent runs. Values less than $\ensuremath{0.001}$ are denoted by $\epsilon$ and greater than $\ensuremath{100.0}$ are denoted by $\Omega$.}
\label{tab:si_anisoap_errors_benzene}
\begin{tabular}{lcccc}
\toprule
 & $(w_E,w_F,w_\tau)$ & $E:\;(\ell^2,\mathrm{RMSE},R^2)$ & $F:\;(\ell^2,\mathrm{RMSE},R^2)$ & $\tau:\;(\ell^2,\mathrm{RMSE},R^2)$ \\
\midrule
\multirow[c]{18}{*}{\makebox[0pt][c]{\rotatebox[origin=c]{90}{\makebox[0pt][c]{\anisoap}}}} & $(0,\,0,\,1)$ & $(1 \pm \epsilon,\,\Omega,\,\ll 0)$ & $(1 \pm \epsilon,\,\Omega,\,\ll 0)$ & $(0.35 \pm 0.016,\,1.1 \pm 0.028,\,0.88 \pm 0.011)$ \\
 & $(0,\,1,\,0)$ & $(0.89 \pm 0.24,\,10 \pm 8.6,\,\ll 0)$ & $(0.2 \pm 0.0099,\,0.57 \pm 0.027,\,0.96 \pm 0.004)$ & $(0.89 \pm 0.24,\,28 \pm 24,\,\ll 0)$ \\
 & $(0,\,\frac{1}{2},\,\frac{1}{2})$ & $(0.8 \pm 0.27,\,0.87 \pm 0.46,\,-0.633 \pm 1.4)$ & $(0.2 \pm 0.0097,\,0.58 \pm 0.029,\,0.959 \pm 0.004)$ & $(0.35 \pm 0.016,\,1.1 \pm 0.029,\,0.88 \pm 0.011)$ \\
 & $(0,\,\frac{1}{3},\,\frac{2}{3})$ & $(0.88 \pm 0.2,\,1.1 \pm 0.5,\,-1.31 \pm 1.7)$ & $(0.2 \pm 0.0097,\,0.58 \pm 0.029,\,0.959 \pm 0.004)$ & $(0.35 \pm 0.016,\,1.1 \pm 0.029,\,0.88 \pm 0.011)$ \\
 & $(0,\,\frac{2}{3},\,\frac{1}{3})$ & $(0.77 \pm 0.31,\,0.73 \pm 0.4,\,-0.171 \pm 1)$ & $(0.2 \pm 0.0096,\,0.58 \pm 0.028,\,0.959 \pm 0.0039)$ & $(0.35 \pm 0.016,\,1.1 \pm 0.029,\,0.879 \pm 0.011)$ \\
 & $(1,\,0,\,0)$ & $(0.17 \pm 0.0098,\,0.13 \pm 0.0065,\,0.97 \pm 0.0034)$ & $(0.27 \pm 0.011,\,0.77 \pm 0.051,\,0.927 \pm 0.0061)$ & $(0.4 \pm 0.011,\,1.3 \pm 0.041,\,0.839 \pm 0.0086)$ \\
 & $(\frac{1}{2},\,0,\,\frac{1}{2})$ & $(0.22 \pm 0.017,\,0.16 \pm 0.01,\,0.953 \pm 0.0076)$ & $(0.21 \pm 0.0077,\,0.6 \pm 0.029,\,0.956 \pm 0.0032)$ & $(0.35 \pm 0.015,\,1.1 \pm 0.028,\,0.88 \pm 0.01)$ \\
 & $(\frac{1}{2},\,\frac{1}{2},\,0)$ & $(0.21 \pm 0.016,\,0.16 \pm 0.0093,\,0.955 \pm 0.0068)$ & $(0.2 \pm 0.0088,\,0.58 \pm 0.027,\,0.959 \pm 0.0036)$ & $(0.36 \pm 0.015,\,1.1 \pm 0.028,\,0.874 \pm 0.011)$ \\
 & $(\frac{1}{2},\,\frac{1}{4},\,\frac{1}{4})$ & $(0.21 \pm 0.017,\,0.16 \pm 0.01,\,0.954 \pm 0.0075)$ & $(0.2 \pm 0.0093,\,0.58 \pm 0.029,\,0.958 \pm 0.0038)$ & $(0.35 \pm 0.015,\,1.1 \pm 0.03,\,0.879 \pm 0.011)$ \\
 & $(\frac{1}{3},\,0,\,\frac{2}{3})$ & $(0.22 \pm 0.018,\,0.17 \pm 0.011,\,0.951 \pm 0.0081)$ & $(0.21 \pm 0.0082,\,0.6 \pm 0.029,\,0.956 \pm 0.0034)$ & $(0.35 \pm 0.015,\,1.1 \pm 0.028,\,0.88 \pm 0.011)$ \\
 & $(\frac{1}{3},\,\frac{1}{3},\,\frac{1}{3})$ & $(0.22 \pm 0.018,\,0.17 \pm 0.011,\,0.951 \pm 0.008)$ & $(0.2 \pm 0.0094,\,0.58 \pm 0.029,\,0.959 \pm 0.0038)$ & $(0.35 \pm 0.016,\,1.1 \pm 0.029,\,0.88 \pm 0.011)$ \\
 & $(\frac{1}{3},\,\frac{2}{3},\,0)$ & $(0.22 \pm 0.017,\,0.17 \pm 0.01,\,0.951 \pm 0.0076)$ & $(0.2 \pm 0.0096,\,0.58 \pm 0.028,\,0.959 \pm 0.0039)$ & $(0.36 \pm 0.015,\,1.1 \pm 0.027,\,0.873 \pm 0.011)$ \\
 & $(\frac{1}{4},\,\frac{1}{2},\,\frac{1}{4})$ & $(0.22 \pm 0.018,\,0.17 \pm 0.011,\,0.95 \pm 0.0083)$ & $(0.2 \pm 0.0095,\,0.58 \pm 0.028,\,0.959 \pm 0.0039)$ & $(0.35 \pm 0.016,\,1.1 \pm 0.03,\,0.879 \pm 0.011)$ \\
 & $(\frac{1}{4},\,\frac{1}{4},\,\frac{1}{2})$ & $(0.22 \pm 0.018,\,0.17 \pm 0.011,\,0.95 \pm 0.0083)$ & $(0.2 \pm 0.0095,\,0.58 \pm 0.029,\,0.959 \pm 0.0039)$ & $(0.35 \pm 0.016,\,1.1 \pm 0.029,\,0.88 \pm 0.011)$ \\
 & $(\frac{2}{3},\,0,\,\frac{1}{3})$ & $(0.21 \pm 0.016,\,0.16 \pm 0.0098,\,0.957 \pm 0.0069)$ & $(0.21 \pm 0.006,\,0.61 \pm 0.026,\,0.955 \pm 0.0026)$ & $(0.35 \pm 0.015,\,1.1 \pm 0.03,\,0.878 \pm 0.01)$ \\
 & $(\frac{2}{3},\,\frac{1}{3},\,0)$ & $(0.2 \pm 0.014,\,0.15 \pm 0.0083,\,0.959 \pm 0.0059)$ & $(0.21 \pm 0.0077,\,0.59 \pm 0.024,\,0.958 \pm 0.0032)$ & $(0.36 \pm 0.014,\,1.1 \pm 0.03,\,0.873 \pm 0.01)$ \\
 & $(\frac{9}{10},\,0.009,\,0.09)$ & $(0.19 \pm 0.014,\,0.14 \pm 0.0079,\,0.965 \pm 0.0052)$ & $(0.22 \pm 0.0075,\,0.64 \pm 0.029,\,0.95 \pm 0.0034)$ & $(0.36 \pm 0.014,\,1.1 \pm 0.03,\,0.87 \pm 0.01)$ \\
 & $(\frac{9}{10},\,0.09,\,0.009)$ & $(0.18 \pm 0.013,\,0.14 \pm 0.0074,\,0.966 \pm 0.0049)$ & $(0.22 \pm 0.0087,\,0.64 \pm 0.03,\,0.95 \pm 0.0038)$ & $(0.37 \pm 0.014,\,1.2 \pm 0.034,\,0.864 \pm 0.01)$ \\
\addlinespace[2pt]
\cmidrule(lr){2-5}
\addlinespace[2pt]
\multirow[c]{18}{*}{\makebox[0pt][c]{\rotatebox[origin=c]{90}{\makebox[0pt][c]{SOAP}}}} & $(0,\,0,\,1)$ & $(1 \pm \epsilon,\,0.75 \pm 0.032,\,\ensuremath{-8.14\times 10^{-5}} \pm \epsilon)$ & $(1 \pm \epsilon,\,2.8 \pm 0.064,\,\ensuremath{-5.55\times 10^{-17}} \pm \epsilon)$ &  \\
 & $(0,\,1,\,0)$ & $(0.99 \pm 0.0089,\,0.77 \pm 0.06,\,-0.0578 \pm 0.099)$ & $(0.71 \pm 0.0091,\,2 \pm 0.068,\,0.494 \pm 0.013)$ &  \\
 & $(0,\,\frac{1}{2},\,\frac{1}{2})$ & $(0.99 \pm 0.016,\,0.77 \pm 0.06,\,-0.047 \pm 0.094)$ & $(0.71 \pm 0.0091,\,2 \pm 0.068,\,0.494 \pm 0.013)$ &  \\
 & $(0,\,\frac{1}{3},\,\frac{2}{3})$ & $(0.99 \pm 0.0089,\,0.77 \pm 0.06,\,-0.0578 \pm 0.099)$ & $(0.71 \pm 0.0091,\,2 \pm 0.068,\,0.494 \pm 0.013)$ &  \\
 & $(0,\,\frac{2}{3},\,\frac{1}{3})$ & $(1 \pm 0.0045,\,0.78 \pm 0.057,\,-0.069 \pm 0.098)$ & $(0.71 \pm 0.0091,\,2 \pm 0.068,\,0.494 \pm 0.013)$ &  \\
 & $(1,\,0,\,0)$ & $(0.81 \pm 0.013,\,0.61 \pm 0.033,\,0.343 \pm 0.021)$ & $(0.8 \pm 0.034,\,2.3 \pm 0.13,\,0.358 \pm 0.055)$ &  \\
 & $(\frac{1}{2},\,0,\,\frac{1}{2})$ & $(0.81 \pm 0.014,\,0.61 \pm 0.033,\,0.343 \pm 0.022)$ & $(0.8 \pm 0.03,\,2.3 \pm 0.12,\,0.363 \pm 0.049)$ &  \\
 & $(\frac{1}{2},\,\frac{1}{2},\,0)$ & $(0.87 \pm 0.02,\,0.66 \pm 0.032,\,0.236 \pm 0.035)$ & $(0.72 \pm 0.01,\,2 \pm 0.074,\,0.482 \pm 0.015)$ &  \\
 & $(\frac{1}{2},\,\frac{1}{4},\,\frac{1}{4})$ & $(0.85 \pm 0.023,\,0.64 \pm 0.034,\,0.274 \pm 0.039)$ & $(0.73 \pm 0.023,\,2.1 \pm 0.1,\,0.465 \pm 0.034)$ &  \\
 & $(\frac{1}{3},\,0,\,\frac{2}{3})$ & $(0.81 \pm 0.013,\,0.61 \pm 0.033,\,0.343 \pm 0.021)$ & $(0.8 \pm 0.034,\,2.3 \pm 0.13,\,0.358 \pm 0.055)$ &  \\
 & $(\frac{1}{3},\,\frac{1}{3},\,\frac{1}{3})$ & $(0.87 \pm 0.02,\,0.66 \pm 0.032,\,0.237 \pm 0.035)$ & $(0.72 \pm 0.012,\,2 \pm 0.077,\,0.481 \pm 0.017)$ &  \\
 & $(\frac{1}{3},\,\frac{2}{3},\,0)$ & $(0.89 \pm 0.019,\,0.67 \pm 0.031,\,0.202 \pm 0.034)$ & $(0.72 \pm 0.0082,\,2 \pm 0.069,\,0.489 \pm 0.012)$ &  \\
 & $(\frac{1}{4},\,\frac{1}{2},\,\frac{1}{4})$ & $(0.89 \pm 0.018,\,0.67 \pm 0.031,\,0.201 \pm 0.033)$ & $(0.71 \pm 0.0082,\,2 \pm 0.068,\,0.49 \pm 0.012)$ &  \\
 & $(\frac{1}{4},\,\frac{1}{4},\,\frac{1}{2})$ & $(0.87 \pm 0.021,\,0.66 \pm 0.032,\,0.237 \pm 0.036)$ & $(0.72 \pm 0.013,\,2 \pm 0.08,\,0.479 \pm 0.019)$ &  \\
 & $(\frac{2}{3},\,0,\,\frac{1}{3})$ & $(0.81 \pm 0.014,\,0.61 \pm 0.034,\,0.344 \pm 0.023)$ & $(0.8 \pm 0.027,\,2.3 \pm 0.12,\,0.364 \pm 0.044)$ &  \\
 & $(\frac{2}{3},\,\frac{1}{3},\,0)$ & $(0.85 \pm 0.022,\,0.64 \pm 0.033,\,0.273 \pm 0.038)$ & $(0.73 \pm 0.02,\,2.1 \pm 0.093,\,0.468 \pm 0.029)$ &  \\
 & $(\frac{9}{10},\,0.009,\,0.09)$ & $(0.81 \pm 0.013,\,0.61 \pm 0.033,\,0.343 \pm 0.022)$ & $(0.79 \pm 0.034,\,2.2 \pm 0.13,\,0.374 \pm 0.055)$ &  \\
 & $(\frac{9}{10},\,0.09,\,0.009)$ & $(0.82 \pm 0.018,\,0.62 \pm 0.033,\,0.328 \pm 0.029)$ & $(0.76 \pm 0.037,\,2.2 \pm 0.13,\,0.42 \pm 0.057)$ &  \\
\bottomrule
\end{tabular}
\end{table}

\begin{table}[p]
\centering
\footnotesize
\setlength{\tabcolsep}{3pt}
\renewcommand{\arraystretch}{1.12}
\caption{Errors for linear SOAP and \anisoap~models for Water on the test set. Loss weights are reported as $(w_E,w_F,w_\tau)$. Each error entry is $(\ell^2,\mathrm{RMSE},R^2)$. Torque errors are reported for \anisoap and omitted for isotropic SOAP. Values are mean $\pm$ standard deviation over equivalent runs. Values less than $\ensuremath{0.001}$ are denoted by $\epsilon$ and greater than $\ensuremath{100.0}$ are denoted by $\Omega$.}
\label{tab:si_anisoap_errors_water}
\begin{tabular}{lcccc}
\toprule
 & $(w_E,w_F,w_\tau)$ & $E:\;(\ell^2,\mathrm{RMSE},R^2)$ & $F:\;(\ell^2,\mathrm{RMSE},R^2)$ & $\tau:\;(\ell^2,\mathrm{RMSE},R^2)$ \\
\midrule
\multirow[c]{18}{*}{\makebox[0pt][c]{\rotatebox[origin=c]{90}{\makebox[0pt][c]{\anisoap}}}} & $(0,\,0,\,1)$ & $(1 \pm \epsilon,\,0.24 \pm 0.024,\,\ll 0)$ & $(1 \pm \epsilon,\,1.4 \pm 0.11,\,\ll 0)$ & $(0.92 \pm 0.0022,\,0.11 \pm 0.0058,\,0.158 \pm 0.004)$ \\
 & $(0,\,1,\,0)$ & $(0.97 \pm 0.067,\,0.09 \pm 0.038,\,-2.37 \pm 2.6)$ & $(0.52 \pm 0.012,\,0.11 \pm 0.0051,\,0.73 \pm 0.012)$ & $(1 \pm \epsilon,\,2.4 \pm 0.57,\,\ll 0)$ \\
 & $(0,\,\frac{1}{2},\,\frac{1}{2})$ & $(0.92 \pm 0.011,\,0.048 \pm 0.0015,\,0.151 \pm 0.021)$ & $(0.52 \pm 0.012,\,0.11 \pm 0.0053,\,0.732 \pm 0.013)$ & $(0.92 \pm 0.0021,\,0.11 \pm 0.0057,\,0.157 \pm 0.0039)$ \\
 & $(0,\,\frac{1}{3},\,\frac{2}{3})$ & $(0.9 \pm 0.027,\,0.047 \pm 0.0024,\,0.191 \pm 0.049)$ & $(0.52 \pm 0.012,\,0.11 \pm 0.0052,\,0.731 \pm 0.013)$ & $(0.92 \pm 0.0024,\,0.11 \pm 0.0057,\,0.157 \pm 0.0043)$ \\
 & $(0,\,\frac{2}{3},\,\frac{1}{3})$ & $(0.94 \pm 0.034,\,0.049 \pm 0.0029,\,0.124 \pm 0.064)$ & $(0.52 \pm 0.012,\,0.11 \pm 0.0053,\,0.731 \pm 0.012)$ & $(0.92 \pm 0.0022,\,0.11 \pm 0.0058,\,0.156 \pm 0.0041)$ \\
 & $(1,\,0,\,0)$ & $(0.84 \pm 0.0035,\,0.044 \pm 0.0012,\,0.292 \pm 0.0058)$ & $(1 \pm \epsilon,\,1.2 \pm 0.99,\,\ll 0)$ & $(1 \pm \epsilon,\,1.4 \pm 1,\,\ll 0)$ \\
 & $(\frac{1}{2},\,0,\,\frac{1}{2})$ & $(0.84 \pm 0.015,\,0.044 \pm 0.0016,\,0.292 \pm 0.025)$ & $(0.53 \pm 0.014,\,0.11 \pm 0.0057,\,0.716 \pm 0.015)$ & $(0.92 \pm 0.0015,\,0.11 \pm 0.0058,\,0.155 \pm 0.0028)$ \\
 & $(\frac{1}{2},\,\frac{1}{2},\,0)$ & $(0.84 \pm 0.013,\,0.044 \pm 0.0015,\,0.291 \pm 0.022)$ & $(0.52 \pm 0.013,\,0.11 \pm 0.0051,\,0.73 \pm 0.014)$ & $(1 \pm \epsilon,\,1.6 \pm 0.47,\,\ll 0)$ \\
 & $(\frac{1}{2},\,\frac{1}{4},\,\frac{1}{4})$ & $(0.84 \pm 0.015,\,0.044 \pm 0.0016,\,0.293 \pm 0.025)$ & $(0.52 \pm 0.013,\,0.11 \pm 0.0053,\,0.73 \pm 0.014)$ & $(0.92 \pm 0.0011,\,0.11 \pm 0.0058,\,0.156 \pm 0.002)$ \\
 & $(\frac{1}{3},\,0,\,\frac{2}{3})$ & $(0.84 \pm 0.016,\,0.044 \pm 0.0017,\,0.291 \pm 0.027)$ & $(0.54 \pm 0.012,\,0.11 \pm 0.006,\,0.711 \pm 0.012)$ & $(0.92 \pm 0.0018,\,0.11 \pm 0.0058,\,0.156 \pm 0.0034)$ \\
 & $(\frac{1}{3},\,\frac{1}{3},\,\frac{1}{3})$ & $(0.84 \pm 0.016,\,0.044 \pm 0.0016,\,0.291 \pm 0.027)$ & $(0.52 \pm 0.013,\,0.11 \pm 0.0053,\,0.731 \pm 0.013)$ & $(0.92 \pm 0.0014,\,0.11 \pm 0.0058,\,0.156 \pm 0.0026)$ \\
 & $(\frac{1}{3},\,\frac{2}{3},\,0)$ & $(0.84 \pm 0.014,\,0.044 \pm 0.0016,\,0.288 \pm 0.024)$ & $(0.52 \pm 0.013,\,0.11 \pm 0.005,\,0.731 \pm 0.013)$ & $(1 \pm \epsilon,\,1.9 \pm 0.52,\,\ll 0)$ \\
 & $(\frac{1}{4},\,\frac{1}{2},\,\frac{1}{4})$ & $(0.84 \pm 0.016,\,0.044 \pm 0.0016,\,0.29 \pm 0.026)$ & $(0.52 \pm 0.012,\,0.11 \pm 0.0053,\,0.731 \pm 0.013)$ & $(0.92 \pm 0.0016,\,0.11 \pm 0.0058,\,0.156 \pm 0.003)$ \\
 & $(\frac{1}{4},\,\frac{1}{4},\,\frac{1}{2})$ & $(0.84 \pm 0.016,\,0.044 \pm 0.0017,\,0.291 \pm 0.028)$ & $(0.52 \pm 0.013,\,0.11 \pm 0.0052,\,0.731 \pm 0.013)$ & $(0.92 \pm 0.0017,\,0.11 \pm 0.0058,\,0.156 \pm 0.0031)$ \\
 & $(\frac{2}{3},\,0,\,\frac{1}{3})$ & $(0.84 \pm 0.013,\,0.044 \pm 0.0015,\,0.293 \pm 0.022)$ & $(0.53 \pm 0.014,\,0.11 \pm 0.0053,\,0.718 \pm 0.015)$ & $(0.92 \pm 0.0013,\,0.11 \pm 0.0058,\,0.153 \pm 0.0024)$ \\
 & $(\frac{2}{3},\,\frac{1}{3},\,0)$ & $(0.84 \pm 0.012,\,0.044 \pm 0.0015,\,0.293 \pm 0.021)$ & $(0.52 \pm 0.013,\,0.11 \pm 0.0052,\,0.729 \pm 0.014)$ & $(1 \pm \epsilon,\,1.4 \pm 0.5,\,\ll 0)$ \\
 & $(\frac{9}{10},\,0.009,\,0.09)$ & $(0.84 \pm 0.0096,\,0.044 \pm 0.0014,\,0.297 \pm 0.016)$ & $(0.53 \pm 0.017,\,0.11 \pm 0.0058,\,0.715 \pm 0.018)$ & $(0.92 \pm 0.0015,\,0.11 \pm 0.0058,\,0.147 \pm 0.0027)$ \\
 & $(\frac{9}{10},\,0.09,\,0.009)$ & $(0.84 \pm 0.009,\,0.044 \pm 0.0014,\,0.297 \pm 0.015)$ & $(0.53 \pm 0.014,\,0.11 \pm 0.0052,\,0.722 \pm 0.015)$ & $(0.93 \pm 0.0022,\,0.11 \pm 0.0059,\,0.14 \pm 0.004)$ \\
\addlinespace[2pt]
\cmidrule(lr){2-5}
\addlinespace[2pt]
\multirow[c]{18}{*}{\makebox[0pt][c]{\rotatebox[origin=c]{90}{\makebox[0pt][c]{SOAP}}}} & $(0,\,0,\,1)$ & $(1 \pm \epsilon,\,0.052 \pm \epsilon,\,\ensuremath{-2.11\times 10^{-5}} \pm \epsilon)$ & $(1 \pm \epsilon,\,0.21 \pm 0.0035,\,\ensuremath{-1.48\times 10^{-16}} \pm \epsilon)$ &  \\
 & $(0,\,1,\,0)$ & $(0.87 \pm 0.012,\,0.045 \pm 0.0014,\,0.243 \pm 0.021)$ & $(0.54 \pm 0.023,\,0.11 \pm 0.0044,\,0.707 \pm 0.026)$ &  \\
 & $(0,\,\frac{1}{2},\,\frac{1}{2})$ & $(0.87 \pm 0.012,\,0.045 \pm 0.0014,\,0.243 \pm 0.021)$ & $(0.54 \pm 0.024,\,0.11 \pm 0.0045,\,0.708 \pm 0.026)$ &  \\
 & $(0,\,\frac{1}{3},\,\frac{2}{3})$ & $(0.87 \pm 0.012,\,0.045 \pm 0.0014,\,0.243 \pm 0.021)$ & $(0.54 \pm 0.024,\,0.11 \pm 0.0045,\,0.708 \pm 0.026)$ &  \\
 & $(0,\,\frac{2}{3},\,\frac{1}{3})$ & $(0.87 \pm 0.012,\,0.045 \pm 0.0014,\,0.243 \pm 0.021)$ & $(0.54 \pm 0.023,\,0.11 \pm 0.0044,\,0.707 \pm 0.025)$ &  \\
 & $(1,\,0,\,0)$ & $(0.87 \pm 0.012,\,0.045 \pm 0.0014,\,0.244 \pm 0.021)$ & $(0.55 \pm 0.023,\,0.12 \pm 0.0048,\,0.692 \pm 0.026)$ &  \\
 & $(\frac{1}{2},\,0,\,\frac{1}{2})$ & $(0.87 \pm 0.012,\,0.045 \pm 0.0014,\,0.244 \pm 0.021)$ & $(0.55 \pm 0.024,\,0.12 \pm 0.005,\,0.693 \pm 0.027)$ &  \\
 & $(\frac{1}{2},\,\frac{1}{2},\,0)$ & $(0.87 \pm 0.012,\,0.045 \pm 0.0015,\,0.245 \pm 0.022)$ & $(0.54 \pm 0.024,\,0.11 \pm 0.0045,\,0.707 \pm 0.026)$ &  \\
 & $(\frac{1}{2},\,\frac{1}{4},\,\frac{1}{4})$ & $(0.87 \pm 0.013,\,0.045 \pm 0.0015,\,0.245 \pm 0.022)$ & $(0.54 \pm 0.024,\,0.11 \pm 0.0045,\,0.707 \pm 0.026)$ &  \\
 & $(\frac{1}{3},\,0,\,\frac{2}{3})$ & $(0.87 \pm 0.012,\,0.045 \pm 0.0014,\,0.244 \pm 0.021)$ & $(0.55 \pm 0.023,\,0.12 \pm 0.0048,\,0.692 \pm 0.026)$ &  \\
 & $(\frac{1}{3},\,\frac{1}{3},\,\frac{1}{3})$ & $(0.87 \pm 0.012,\,0.045 \pm 0.0015,\,0.245 \pm 0.022)$ & $(0.54 \pm 0.023,\,0.11 \pm 0.0044,\,0.707 \pm 0.026)$ &  \\
 & $(\frac{1}{3},\,\frac{2}{3},\,0)$ & $(0.87 \pm 0.012,\,0.045 \pm 0.0014,\,0.245 \pm 0.022)$ & $(0.54 \pm 0.024,\,0.11 \pm 0.0044,\,0.707 \pm 0.026)$ &  \\
 & $(\frac{1}{4},\,\frac{1}{2},\,\frac{1}{4})$ & $(0.87 \pm 0.012,\,0.045 \pm 0.0014,\,0.245 \pm 0.022)$ & $(0.54 \pm 0.023,\,0.11 \pm 0.0044,\,0.707 \pm 0.026)$ &  \\
 & $(\frac{1}{4},\,\frac{1}{4},\,\frac{1}{2})$ & $(0.87 \pm 0.012,\,0.045 \pm 0.0015,\,0.245 \pm 0.022)$ & $(0.54 \pm 0.023,\,0.11 \pm 0.0044,\,0.707 \pm 0.025)$ &  \\
 & $(\frac{2}{3},\,0,\,\frac{1}{3})$ & $(0.87 \pm 0.012,\,0.045 \pm 0.0014,\,0.243 \pm 0.02)$ & $(0.56 \pm 0.021,\,0.12 \pm 0.0047,\,0.691 \pm 0.024)$ &  \\
 & $(\frac{2}{3},\,\frac{1}{3},\,0)$ & $(0.87 \pm 0.013,\,0.045 \pm 0.0015,\,0.245 \pm 0.022)$ & $(0.54 \pm 0.024,\,0.11 \pm 0.0045,\,0.707 \pm 0.026)$ &  \\
 & $(\frac{9}{10},\,0.009,\,0.09)$ & $(0.87 \pm 0.012,\,0.045 \pm 0.0014,\,0.245 \pm 0.022)$ & $(0.55 \pm 0.023,\,0.11 \pm 0.0046,\,0.698 \pm 0.026)$ &  \\
 & $(\frac{9}{10},\,0.09,\,0.009)$ & $(0.87 \pm 0.013,\,0.045 \pm 0.0015,\,0.246 \pm 0.022)$ & $(0.54 \pm 0.025,\,0.11 \pm 0.0046,\,0.705 \pm 0.027)$ &  \\
\bottomrule
\end{tabular}
\end{table}

\begin{table}[p]
\centering
\footnotesize
\setlength{\tabcolsep}{3pt}
\renewcommand{\arraystretch}{1.12}
\caption{Errors for linear SOAP and \anisoap~models for Formamide on the test set. Loss weights are reported as $(w_E,w_F,w_\tau)$. Each error entry is $(\ell^2,\mathrm{RMSE},R^2)$. Torque errors are reported for \anisoap and omitted for isotropic SOAP. Values are mean $\pm$ standard deviation over equivalent runs. Values less than $\ensuremath{0.001}$ are denoted by $\epsilon$ and greater than $\ensuremath{100.0}$ are denoted by $\Omega$.}
\label{tab:si_anisoap_errors_formamide}
\begin{tabular}{lcccc}
\toprule
 & $(w_E,w_F,w_\tau)$ & $E:\;(\ell^2,\mathrm{RMSE},R^2)$ & $F:\;(\ell^2,\mathrm{RMSE},R^2)$ & $\tau:\;(\ell^2,\mathrm{RMSE},R^2)$ \\
\midrule
\multirow[c]{18}{*}{\makebox[0pt][c]{\rotatebox[origin=c]{90}{\makebox[0pt][c]{\anisoap}}}} & $(0,\,0,\,1)$ & $(1 \pm \epsilon,\,9.5 \pm 5.4,\,\ll 0)$ & $(1 \pm \epsilon,\,22 \pm 12,\,\ll 0)$ & $(0.67 \pm 0.0097,\,0.5 \pm 0.032,\,0.558 \pm 0.013)$ \\
 & $(0,\,1,\,0)$ & $(0.95 \pm 0.086,\,0.56 \pm 0.63,\,\ll 0)$ & $(0.51 \pm 0.011,\,0.46 \pm 0.025,\,0.74 \pm 0.011)$ & $(0.85 \pm 0.17,\,1.5 \pm 1.6,\,-5.99 \pm 11)$ \\
 & $(0,\,\frac{1}{2},\,\frac{1}{2})$ & $(0.95 \pm 0.082,\,0.22 \pm 0.082,\,-0.472 \pm 1.1)$ & $(0.51 \pm 0.011,\,0.46 \pm 0.025,\,0.739 \pm 0.011)$ & $(0.67 \pm 0.009,\,0.5 \pm 0.032,\,0.557 \pm 0.012)$ \\
 & $(0,\,\frac{1}{3},\,\frac{2}{3})$ & $(1 \pm \epsilon,\,0.25 \pm 0.046,\,-0.701 \pm 0.51)$ & $(0.51 \pm 0.011,\,0.46 \pm 0.025,\,0.738 \pm 0.011)$ & $(0.66 \pm 0.009,\,0.5 \pm 0.031,\,0.558 \pm 0.012)$ \\
 & $(0,\,\frac{2}{3},\,\frac{1}{3})$ & $(0.94 \pm 0.11,\,0.24 \pm 0.11,\,-0.763 \pm 1.5)$ & $(0.51 \pm 0.0098,\,0.46 \pm 0.024,\,0.739 \pm 0.01)$ & $(0.67 \pm 0.0088,\,0.5 \pm 0.031,\,0.555 \pm 0.012)$ \\
 & $(1,\,0,\,0)$ & $(0.64 \pm 0.012,\,0.12 \pm 0.0054,\,0.593 \pm 0.015)$ & $(0.63 \pm 0.015,\,0.57 \pm 0.029,\,0.598 \pm 0.019)$ & $(0.75 \pm 0.019,\,0.56 \pm 0.043,\,0.443 \pm 0.028)$ \\
 & $(\frac{1}{2},\,0,\,\frac{1}{2})$ & $(0.68 \pm 0.019,\,0.13 \pm 0.0032,\,0.537 \pm 0.026)$ & $(0.53 \pm 0.014,\,0.48 \pm 0.028,\,0.714 \pm 0.015)$ & $(0.67 \pm 0.013,\,0.5 \pm 0.035,\,0.551 \pm 0.018)$ \\
 & $(\frac{1}{2},\,\frac{1}{2},\,0)$ & $(0.69 \pm 0.018,\,0.13 \pm 0.0042,\,0.53 \pm 0.025)$ & $(0.52 \pm 0.015,\,0.47 \pm 0.029,\,0.729 \pm 0.016)$ & $(0.69 \pm 0.013,\,0.52 \pm 0.036,\,0.53 \pm 0.018)$ \\
 & $(\frac{1}{2},\,\frac{1}{4},\,\frac{1}{4})$ & $(0.68 \pm 0.017,\,0.13 \pm 0.0034,\,0.532 \pm 0.024)$ & $(0.52 \pm 0.016,\,0.47 \pm 0.029,\,0.728 \pm 0.016)$ & $(0.67 \pm 0.013,\,0.51 \pm 0.035,\,0.548 \pm 0.017)$ \\
 & $(\frac{1}{3},\,0,\,\frac{2}{3})$ & $(0.69 \pm 0.021,\,0.13 \pm 0.003,\,0.52 \pm 0.028)$ & $(0.53 \pm 0.014,\,0.48 \pm 0.028,\,0.719 \pm 0.015)$ & $(0.67 \pm 0.012,\,0.5 \pm 0.034,\,0.554 \pm 0.016)$ \\
 & $(\frac{1}{3},\,\frac{1}{3},\,\frac{1}{3})$ & $(0.7 \pm 0.019,\,0.13 \pm 0.0032,\,0.514 \pm 0.026)$ & $(0.52 \pm 0.014,\,0.47 \pm 0.028,\,0.733 \pm 0.015)$ & $(0.67 \pm 0.012,\,0.5 \pm 0.034,\,0.553 \pm 0.015)$ \\
 & $(\frac{1}{3},\,\frac{2}{3},\,0)$ & $(0.7 \pm 0.02,\,0.13 \pm 0.0043,\,0.513 \pm 0.027)$ & $(0.52 \pm 0.013,\,0.47 \pm 0.027,\,0.734 \pm 0.014)$ & $(0.68 \pm 0.014,\,0.51 \pm 0.036,\,0.536 \pm 0.019)$ \\
 & $(\frac{1}{4},\,\frac{1}{2},\,\frac{1}{4})$ & $(0.7 \pm 0.019,\,0.13 \pm 0.0032,\,0.505 \pm 0.027)$ & $(0.51 \pm 0.014,\,0.46 \pm 0.028,\,0.736 \pm 0.015)$ & $(0.67 \pm 0.011,\,0.5 \pm 0.034,\,0.552 \pm 0.015)$ \\
 & $(\frac{1}{4},\,\frac{1}{4},\,\frac{1}{2})$ & $(0.7 \pm 0.02,\,0.13 \pm 0.003,\,0.506 \pm 0.028)$ & $(0.52 \pm 0.014,\,0.47 \pm 0.028,\,0.734 \pm 0.014)$ & $(0.67 \pm 0.011,\,0.5 \pm 0.034,\,0.555 \pm 0.015)$ \\
 & $(\frac{2}{3},\,0,\,\frac{1}{3})$ & $(0.67 \pm 0.018,\,0.13 \pm 0.0035,\,0.557 \pm 0.023)$ & $(0.54 \pm 0.014,\,0.49 \pm 0.028,\,0.703 \pm 0.015)$ & $(0.68 \pm 0.015,\,0.51 \pm 0.037,\,0.543 \pm 0.02)$ \\
 & $(\frac{2}{3},\,\frac{1}{3},\,0)$ & $(0.67 \pm 0.016,\,0.13 \pm 0.0043,\,0.548 \pm 0.022)$ & $(0.53 \pm 0.015,\,0.48 \pm 0.029,\,0.721 \pm 0.016)$ & $(0.69 \pm 0.013,\,0.52 \pm 0.036,\,0.52 \pm 0.018)$ \\
 & $(\frac{9}{10},\,0.009,\,0.09)$ & $(0.64 \pm 0.014,\,0.12 \pm 0.0045,\,0.588 \pm 0.018)$ & $(0.58 \pm 0.014,\,0.52 \pm 0.03,\,0.668 \pm 0.016)$ & $(0.7 \pm 0.018,\,0.53 \pm 0.04,\,0.506 \pm 0.025)$ \\
 & $(\frac{9}{10},\,0.09,\,0.009)$ & $(0.65 \pm 0.012,\,0.12 \pm 0.0046,\,0.577 \pm 0.016)$ & $(0.56 \pm 0.016,\,0.51 \pm 0.03,\,0.687 \pm 0.018)$ & $(0.71 \pm 0.016,\,0.54 \pm 0.039,\,0.49 \pm 0.022)$ \\
\addlinespace[2pt]
\cmidrule(lr){2-5}
\addlinespace[2pt]
\multirow[c]{18}{*}{\makebox[0pt][c]{\rotatebox[origin=c]{90}{\makebox[0pt][c]{SOAP}}}} & $(0,\,0,\,1)$ & $(1 \pm \epsilon,\,0.19 \pm 0.014,\,-0.000537 \pm \epsilon)$ & $(1 \pm \epsilon,\,0.91 \pm 0.062,\,\ensuremath{2.78\times 10^{-17}} \pm \epsilon)$ &  \\
 & $(0,\,1,\,0)$ & $(1 \pm \epsilon,\,0.22 \pm 0.054,\,-0.318 \pm 0.52)$ & $(0.82 \pm 0.0083,\,0.75 \pm 0.054,\,0.334 \pm 0.013)$ &  \\
 & $(0,\,\frac{1}{2},\,\frac{1}{2})$ & $(1 \pm \epsilon,\,0.2 \pm 0.016,\,-0.075 \pm 0.056)$ & $(0.82 \pm 0.0083,\,0.75 \pm 0.054,\,0.334 \pm 0.014)$ &  \\
 & $(0,\,\frac{1}{3},\,\frac{2}{3})$ & $(1 \pm \epsilon,\,0.22 \pm 0.054,\,-0.319 \pm 0.52)$ & $(0.82 \pm 0.0083,\,0.75 \pm 0.054,\,0.334 \pm 0.013)$ &  \\
 & $(0,\,\frac{2}{3},\,\frac{1}{3})$ & $(1 \pm \epsilon,\,0.2 \pm 0.026,\,-0.148 \pm 0.16)$ & $(0.82 \pm 0.0083,\,0.75 \pm 0.054,\,0.334 \pm 0.014)$ &  \\
 & $(1,\,0,\,0)$ & $(0.94 \pm 0.0064,\,0.18 \pm 0.012,\,0.121 \pm 0.012)$ & $(0.9 \pm 0.073,\,0.83 \pm 0.13,\,0.186 \pm 0.14)$ &  \\
 & $(\frac{1}{2},\,0,\,\frac{1}{2})$ & $(0.94 \pm 0.0063,\,0.18 \pm 0.012,\,0.12 \pm 0.012)$ & $(0.9 \pm 0.074,\,0.82 \pm 0.13,\,0.19 \pm 0.14)$ &  \\
 & $(\frac{1}{2},\,\frac{1}{2},\,0)$ & $(0.96 \pm 0.016,\,0.18 \pm 0.011,\,0.0761 \pm 0.031)$ & $(0.82 \pm 0.0091,\,0.75 \pm 0.054,\,0.332 \pm 0.015)$ &  \\
 & $(\frac{1}{2},\,\frac{1}{4},\,\frac{1}{4})$ & $(0.96 \pm 0.015,\,0.18 \pm 0.011,\,0.0854 \pm 0.029)$ & $(0.82 \pm 0.0093,\,0.75 \pm 0.055,\,0.331 \pm 0.015)$ &  \\
 & $(\frac{1}{3},\,0,\,\frac{2}{3})$ & $(0.94 \pm 0.0062,\,0.18 \pm 0.012,\,0.121 \pm 0.012)$ & $(0.9 \pm 0.073,\,0.82 \pm 0.12,\,0.188 \pm 0.14)$ &  \\
 & $(\frac{1}{3},\,\frac{1}{3},\,\frac{1}{3})$ & $(0.96 \pm 0.016,\,0.18 \pm 0.011,\,0.0762 \pm 0.031)$ & $(0.82 \pm 0.0091,\,0.75 \pm 0.054,\,0.332 \pm 0.015)$ &  \\
 & $(\frac{1}{3},\,\frac{2}{3},\,0)$ & $(0.97 \pm 0.017,\,0.18 \pm 0.011,\,0.0675 \pm 0.033)$ & $(0.82 \pm 0.0089,\,0.75 \pm 0.054,\,0.333 \pm 0.014)$ &  \\
 & $(\frac{1}{4},\,\frac{1}{2},\,\frac{1}{4})$ & $(0.97 \pm 0.017,\,0.18 \pm 0.011,\,0.0675 \pm 0.033)$ & $(0.82 \pm 0.0089,\,0.75 \pm 0.054,\,0.333 \pm 0.015)$ &  \\
 & $(\frac{1}{4},\,\frac{1}{4},\,\frac{1}{2})$ & $(0.96 \pm 0.016,\,0.18 \pm 0.011,\,0.0762 \pm 0.031)$ & $(0.82 \pm 0.0091,\,0.75 \pm 0.054,\,0.332 \pm 0.015)$ &  \\
 & $(\frac{2}{3},\,0,\,\frac{1}{3})$ & $(0.94 \pm 0.0063,\,0.18 \pm 0.012,\,0.121 \pm 0.012)$ & $(0.9 \pm 0.073,\,0.83 \pm 0.12,\,0.187 \pm 0.14)$ &  \\
 & $(\frac{2}{3},\,\frac{1}{3},\,0)$ & $(0.96 \pm 0.015,\,0.18 \pm 0.011,\,0.0855 \pm 0.029)$ & $(0.82 \pm 0.0092,\,0.75 \pm 0.054,\,0.331 \pm 0.015)$ &  \\
 & $(\frac{9}{10},\,0.009,\,0.09)$ & $(0.94 \pm 0.0076,\,0.18 \pm 0.012,\,0.118 \pm 0.014)$ & $(0.85 \pm 0.024,\,0.78 \pm 0.075,\,0.276 \pm 0.041)$ &  \\
 & $(\frac{9}{10},\,0.09,\,0.009)$ & $(0.95 \pm 0.013,\,0.18 \pm 0.011,\,0.105 \pm 0.024)$ & $(0.82 \pm 0.0095,\,0.75 \pm 0.056,\,0.321 \pm 0.016)$ &  \\
\bottomrule
\end{tabular}
\end{table}

\end{document}